\documentclass[reprint,amsmath,amssymb,aps,pra,floatfix]{revtex4-2}
\usepackage[utf8]{inputenc}
\usepackage{graphicx}
\usepackage{amsfonts}
\usepackage{amsthm}
\usepackage{physics}
\usepackage{bbold}
\usepackage[dvipsnames,usenames]{xcolor}
\usepackage{mathrsfs}
\usepackage{tikz} 
\usetikzlibrary{quantikz}
\usepackage{apptools}
\usepackage{subfig}
\usepackage{multirow}
\usepackage[dvipsnames]{xcolor}
\AtAppendix{\counterwithin{theorem}{section}}
\usepackage{bm}
\usepackage{comment}
\usepackage[colorlinks=true, linkcolor=blue, citecolor=green]{hyperref}

\newcommand{\QFT}{\text{QFT}}
\newcommand{\SUMgate}{\text{SUM}}
\newcommand{\Xd}{\mathcal{X}}
\newcommand{\Zd}{\mathcal{Z}}
\newcommand{\Sd}{\mathcal{S}}
\newcommand{\Td}{\mathcal{T}}

\usepackage{tikz}
\usetikzlibrary{quantikz}

\begin{document}
\UseRawInputEncoding

\title{Qudit stabilizer codes for $\mathbb{Z}_N$ lattice gauge theories with matter}

\newcommand{\UNITN}{{Dipartimento di Fisica, University of Trento, via Sommarive 14, I–38123, Povo, Trento, Italy}}
\newcommand{\TIFPA}{INFN-TIFPA Trento Institute of Fundamental Physics and Applications,  Trento, Italy}
\newcommand{\UT}{{Department of Computer Science, University of Toronto, Toronto, ON M5S 2E4, Canada}}
\newcommand{\PNNL}{{Pacific Northwest National Laboratory, Richland, WA 99354, USA}}

\author{Luca~Spagnoli}
\email{luca.spagnoli@unitn.it}
\affiliation{\UNITN}
\affiliation{\TIFPA}
\author{Alessandro~Roggero}
\email{a.roggero@unitn.it}
\affiliation{\UNITN}
\affiliation{\TIFPA}

\author{Nathan~Wiebe}
\email{nawiebe@cs.toronto.edu}
\affiliation{\UT}
\affiliation{\PNNL}

\date{\today}

%\keywords{TBD}
\begin{abstract}
In this work we extend the connection between Quantum Error Correction (QEC) and Lattice Gauge Theories (LGTs) by showing that a $\mathbb{Z}_N$ gauge theory with prime dimension $N$ coupled to dynamical matter can be expressed as a qudit stabilizer code. Using the stabilizer formalism we show how to formulate an exact mapping of the encoded $\mathbb{Z}_N$ gauge theory onto two different bosonic models, uncovering a logical duality generated by error correction itself. From this perspective, quantum error correction provides a unifying language to expose dual descriptions of lattice gauge theories. In addition, we generalize earlier $\mathbb{Z}_2$ constructions on qubits to $\mathbb{Z}_N$ on $N$-level qudits and demonstrate how universal fault-tolerant gates can be implemented via state injection between compatible qudit codes.
\end{abstract}
\maketitle

\section{Introduction}

Quantum computers are rapidly maturing into platforms capable of tackling problems that are intractable for classical devices. Among the most promising targets for an efficient quantum simulation are lattice gauge theories (LGTs), whose non-perturbative dynamics underlie phenomena from quark confinement in quantum chromodynamics to exotic quantum phases of matter. In the past years, many attempts to find efficient algorithms for the digital quantum simulation of LGTs have been made (for recent reviews see~\cite{Ba_uls_2020, Zohar_2021, Klco_2022, Bauer_PRX_2023}).

Yet algorithmic advances alone cannot overcome the reliability challenge. Even in a near-fault-tolerant regime, where logical error rates are low but not yet negligible, decoherence and gate imperfections accumulate over the long circuit depths required for lattice-gauge theories simulations. Controlling these errors is therefore indispensable, and due to hardware limitations it is essential to find ways to control errors in the most memory-efficient way possible. Quantum-error-correction (QEC)~\cite{ShorQEC,Gottesman1997, Gottesman1998, Knill_qec_98,Terhal2015,Roffe2019} has made steady progress, from surface-code demonstrations below threshold to qudit stabiliser schemes that exploit multi-level structures to cut overhead~\cite{google_beyond_qec_2024, Brock_2025, Gupta_2024}.

Recently, numerous proposals for symmetry-aware error correcting and detecting codes have been studied~\cite{Rajput2023, spagnoli2024faulttolerantsimulationlatticegauge, wauters2024symmetryprotection, Stryker2019, Raychowdhury2020, Faist2020, Klco2021_qec, Kong2022, delPino2023, gustafson2023robustness, Chen2024, carena2024quantum} with recent work extending these constructions to include continuous non-abelian groups in an irrep basis~\cite{yao2025quantumerrorcorrectioncodes}. These strategies enforce local symmetry (or gauge) constraints, to detect or correct errors that would bring the simulation in non-physical sectors of the Hilbert space, while exploiting the redundancy of the gauge symmetry. This work aims to generalize those results, embedding QEC into the algorithmic fabric to reduce simulation resources.  Early proposals have shown that embedding gauge symmetry directly into the computational Hilbert space can reduce the resource costs of such fault-tolerant simulations~\cite{Rajput2023,spagnoli2024faulttolerantsimulationlatticegauge,yao2025quantumerrorcorrectioncodes}. In the recent Gauss's law stabiliser code for $\mathbb{Z}_2$ lattice gauge theories~\cite{spagnoli2024faulttolerantsimulationlatticegauge}, quantum error correction and gauge invariance are combined to achieve fault-tolerant Hamiltonian simulation while eliminating redundant degrees of freedom.

Progress towards more realistic models, however, demands moving beyond the binary variables of $\mathbb{Z}_2$ to higher-dimensional Abelian groups $\mathbb{Z}_N$. These discrete groups interpolate between $\mathbb{Z}_2$ and the continuous $U(1)$ limit providing controlled approximations to quantum electrodynamics~\cite{Kuhn2014,Haase2021resourceefficient}, match the natural Hilbert-space structure offered by emerging qudit hardware such as superconducting platforms and trapped-ion chains~\cite{ciavarella2022conceptualaspectsoperatordesign,PhysRevResearch.6.013202,PRXQuantum.5.040309,kurkoglu2024quditgatedecompositiondependence,jiang2025nonabeliandynamicscubeimproving,goss2022high,ringbauer2022universal,PhysRevX.11.021010} and more generally can show an interesting phase-diagram~\cite{PhysRevD.98.074503,roy2023mathbbznlatticegaugetheories,PhysRevResearch.3.013133}. At the same time, the theoretical foundations for error correction with qudits are now well established: the generalized Pauli group, symplectic stabiliser formalism and associated code-classification tools have been worked out for prime dimensions \cite{jain2020normalformsinglequditcliffordt, Wang_2020, GLAUDELL201954, Gottesman1999, Gheorghiu_2014, Durso-Sabina}. What has been missing is a unifying framework that harnesses these qudit techniques to realise fault-tolerant $\mathbb{Z}_N$ gauge simulations with the same economy, and locality, achieved previously for $\mathbb{Z}_2$ with qubits.

\smallskip

This work fills that gap. We generalize the gauge-covariant coding paradigm to arbitrary $\mathbb{Z}_N$ (with $N$ prime) lattice gauge theories in any spatial dimension, exploiting the multilevel structure of qudits to keep the code local and low-overhead. Dynamical fermions are introduced using a generalization of the Jordan-Wigner encoding to qudit systems introducing non-local terms in the Hamiltonian in more than one spatial dimension. %The key advance is the concept of logical duality that “bosonises” the encoded Hamiltonian: after enforcing Gauss's law as a stabiliser condition we identify a complete set of logical operators that map the gauge theory onto a system of bosons living on the logical qudits.
The mapping preserves the reduction in physical degrees of freedom that characterized the $\mathbb{Z}_2$ case, while accommodating the richer dynamics of higher-dimensional clock variables. Remarkably, the efficiency of the reduction depends on the matter fields on the lattice: for spinless-fermions on the matter sites the logical Hamiltonian still maintains a local $\mathbb{Z}_2$ gauge symmetry which is instead broken if one considers bosonic $N$-level matter on sites. 

By showing that gauge-covariant error correction and logical duality extend seamlessly from qubits to qudits lays the groundwork for future generalizations to more complicated symmetry groups such as non-Abelian groups.

\smallskip

This article is structured to guide the reader from the essential background through to the practical implementation of our results. Section~\ref{section:2} reviews the qudit stabiliser formalism, introducing generalized Pauli operators, Clifford and non-Clifford gates and the error-correction framework that underpins the rest of the work. Section~\ref{section:3} turns to lattice gauge theory itself: we formulate the $\mathbb{Z}_N$ Hamiltonian including fermionic matter, electric and magnetic terms, derive the associated Gauss's law operator, and express everything in the generalized Pauli basis by generalizing the Jordan-Wigner encoding. Section~\ref{section:4} constructs the Gauss's law error-correcting code, identifies its logical operators and rewrites the physical Hamiltonian entirely in terms of those logical degrees of freedom. Building on this, Section~\ref{section:5} establishes the logical duality (our “bosonisation” step), mapping the encoded gauge model onto an interacting hardcore-boson Hamiltonian while preserving the reduction in degrees of freedom achieved for $\mathbb{Z}_2$. Section~\ref{section:universal_gate_set} completes the framework by detailing a universal, fault-tolerant gate set realised through state-injection between compatible qudit codes, enabling the digital quantum simulation of the theory.

\section{Stabilizer formalism for Qudits}
\label{section:2}

In this section we will introduce the formalism of quantum error-correction with qudits, generalizing the construction for qubit stabilizer error-correcting codes. We include a more detailed explanation of the formalism and choices done in generalizing the formalism, as well as an example of phase-flip qudit code in Appendix \ref{Appendix:qudits_error_correction}.

Consider $N$-level qudits, where we assume $N$ to be a prime number so that the group structure of the Pauli and Stabilizer groups are preserved. Let $\Xd$ and $\Zd$ be the generalized Pauli matrices defined as:
\begin{equation}
\begin{split}
    \Xd\ket{j} &= \ket{(j+1) \mod N} \\
    \Zd\ket{j} &= \omega^j \ket{j} \\
\end{split}
\end{equation}
where $\omega = e^{i2\pi/N}$ is the $N$-th root of unity, and $j\in \mathbb{Z}_N$. These generalized Pauli matrices satisfy the following relation
\begin{equation}
    \Xd\Zd = \omega^{-1}\Zd\Xd
\end{equation}
and they generate the generalized single-qudit Pauli group. By taking the tensor product of Pauli matrices over $n$ different qudits we generate the generalized $n$-qudits Pauli group $P_n$. Then, we can define the Stabilizer group $S$ of an error-correcting code as an Abelian subgroup of the generalized Pauli group. The code space is then the set of states that are eigenvectors of all elements of $S$ with eigenvalue $+1$. As in the qubit case, if the stabilizer code has $n$ physical qudits and $n-k$ stabilizers (generators of $S$), then it can encode $k$ logical qudits. We can extend $S$ to a set of $n$ commuting operators, by choosing $k$ additional operators $\bar \Zd_1, \cdots, \bar \Zd_k$ that will play the role of logical $\Zd$ operators on the $k$ encoded qudits. Finally, we can introduce the logical $\Xd$ operators by defining $\bar \Xd_1, \cdots, \bar \Xd_k$ such that every $\bar \Xd_i$ commutes with every element of $S$ and satisfy the following relations:
\begin{equation}
\begin{split}
    \bar \Xd_i \bar \Zd_j &= \bar \Zd_j \bar \Xd_i \quad i\ne j \\
    \bar \Xd_i \bar \Zd_i &= \omega^{-1} \bar \Zd_i \bar \Xd_i\label{eq:pauliCommutation}
\end{split}
\end{equation}
The generators of $S$ plus the logical operations are a set of $n+k$ operators that generate the normalizer $N(S)$ of $S$, and the distance $D$ of the code is the minimum Hamming weight that an operator in $N(S)\setminus S$ can have. We will denote such a code with $[[n, k, d]]_N$. We will call a detectable error, any operator $E$ that satisfies $EM=\omega^a ME$ with $a\ne 0 \mod N$, for at least $1$ operator $M$ in the stabilizer $S$. In general, we may distinguish between the ability of a code to correct $X$ (bit-flip) errors and $Z$ (phase-flip) errors. By denoting with $d_x$ ($d_z$) the minimum Hamming weight of $Z$ ($X$) operators in $N(S)\setminus S$, we will say that a quantum code $[[n,k,\min(d_x,d_z)]]_N$ is also a classical code $[n,k,d_x]_N$ ($[n,k,d_z]_N$). In particular, a classical code, denoted with single square-brackets, is an error-correcting code that aims to correct only one type of error. For example, if $d_x=3$ and $d_z=1$, the quantum code would have $d=1$, but it would still be able to correct any single-qubit bit-flip error. Any classical code can be concatenated with another classical code to build a code with higher distance in one, or both errors.

In analogy with the qubit case, we can also define the Clifford group as:
\begin{equation}
     \mathcal{C} = \left\{ U\in SU(N) : UMU^\dag \in \omega^k P_n, \forall M \in P_n \right\}
\end{equation}
where $P_n$ is the generalized $n$-qudit Pauli group, and $k$ integer. Let us define the following gates:
\begin{equation}
    \QFT = \frac{1}{\sqrt{N}} \sum_{k,j=0}^{N-1} \omega^{jk} \ket{k}\bra{j}
\end{equation}
\begin{equation}
    \Sd = \sum_{j=0}^{N-1} \omega^{j(j+1)[2^{-1}]_N} \ket{j}\bra{j}
\end{equation}
where $\QFT \Zd \QFT^\dagger = \Xd$, and $[2^{-1}]_N$ is the inverse of $2$ in the $\mathbb{Z}_N$ cyclic group (for example, for $N=5$, $[2^{-1}]_5=3$). In Appendix \ref{App:cliffordST} we provide the proof that $\Sd$ as we defined it is Clifford, while one can find the proof that those operators generate the entire single-qudit Clifford group in \cite{jain2020normalformsinglequditcliffordt, Wang_2020, GLAUDELL201954}. Then, as in the qubit case, we need only one additional operator, which is not Clifford, in order to generate a universal set of single-qudit gates. In this paper we will consider the following qudit $\Td$ gate:
\begin{equation}
    \Td = \sum_{j=0}^{N-1} \omega^{j^3 [6^{-1}]_N} \ket{j}\bra{j}
    \label{eq:definition_T}
\end{equation}
To define a universal set of multi-qudit gates, we can add the control-$\Zd$ operator defined as:
\begin{equation}
    C\Zd = \sum_j \ket{j}\bra{j}\otimes Z^j
\end{equation}
and the $\SUMgate$ gate as $\SUMgate = (\mathbb{1}\otimes\QFT)C\Zd (\mathbb{1}\otimes\QFT^\dagger)$. Further details on those gates, examples of how to use them to extract syndromes, as well as an example of error-correcting code, are in Appendix \ref{Appendix:qudits_error_correction}.

\section{Abelian lattice gauge theories}
\label{section:3}

Now that we have reviewed the formalism of qudit quantum error correcting codes, we can use this language in order to formulate Abelian lattice gauge theories. We show how all the ingredients of the $\mathbb{Z}_N$ gauge theory can be translated into the language of qudit operators, providing an explicit mapping and encoding that makes the theory ready for implementation on multilevel quantum hardware. In particular, we will start from the gauge (bosonic) and the fermionic operators, finishing with the Hamiltonian and the Gauss's law operator.

\subsection{From free-fermions to gauge theories}

We start from the simplest model of fermions on a lattice with Hamiltonian given by~\cite{PhysRevD.16.3031,PhysRevD.11.395,Wiese2013}
\begin{equation}
H_{free} = \epsilon\sum_{\langle x,y\rangle} s_{xy}\left(\psi^\dagger_x\psi_y+\psi^\dagger_y\psi_x\right)+m\sum_x s_x\psi^\dagger_x\psi_x\;,
\label{eq:free_fermionic_hamiltonian}
\end{equation}
where $\psi^\dagger_x$ and $\psi_x$ are anti-commuting creation and annihilation operators on site $x$
\begin{equation*}
\{\psi^\dagger_x,\psi_y\}=\delta_{xy}\quad\{\psi^\dagger_x,\psi^\dagger_y\}=\{\psi_x,\psi_y\}=0\;.
\end{equation*}
The parameters $t$ and $m$ control the strength of the hopping term and the mass contribution; we use $\langle x,y\rangle$ to denote nearest neighboring sites on a $D$ dimensional lattice and the coefficients $s_x$ and $s_{xy}$ are position dependent phases used to represent staggered fermions on the lattice in order to avoid the fermion doubling problem~\cite{PhysRevD.16.3031,NIELSEN198120} while reproducing the correct Dirac equation in the continuum limit. With the staggering prescription, even sites of the lattice are associated with the positive energy components of the Dirac spinor (related to particles) while the odd sites are used for the negative energy components (related to anti-particles). The common convention is for the vacuum state to correspond to all the even sites being empty, which can be described with the $\ket{0}$ state on the site, while all the odd sites are filled, which in turn can be described with the $\ket{1}$ state on the site. See Table~\ref{tab:occupation_staggered}. Then, by denoting with $\ket{\Omega}$ this vacuum state, the creation of a particle will be described by the application of $\psi^\dagger_x$ on the vacuum for even $x$, while the creation of an anti-particle will be described as the application of $\psi_x$ for odd $x$ (which corresponds to the creation of a hole in the Dirac-sea that describes the vacuum).

\begin{table}[t]
    \centering
    \begin{tabular}{|c|c|c|}
    \hline
         & even & odd \\
         \hline
        empty & $\ket{0}$ & $\ket{1}$ \\
        \hline
        full & $\ket{1}$ & $\ket{0}$ \\
        \hline
    \end{tabular}
    \caption{Occupation in the staggered fermions representation: the state $\ket{1}$ on even sites means that a fermion is present, while the state $\ket{0}$ on odd sites means that an anti-fermion is present.}
    \label{tab:occupation_staggered}
\end{table}

Let's consider now the action of element of a group $G$ on the system. A common choice is to take fermions to transform according to some faithful representation of the group. For simplicity, here we consider the group to be $\mathbb{Z}_N$, so that the faithful representation is also irreducible (irrep), and thus one-dimensional. This means that there are unitary operators $\mathcal{U}_x(g)$ for every element $g\in G$ acting on the fermionic operators on the site $x$ as follows
\begin{equation}
\mathcal{U}_x(g)\psi_x^\dagger \mathcal{U}^\dagger_x(g)=\mathcal{D}(g)\psi_x^\dagger
\label{eq:psi_dag_transformation}
\end{equation}
with $\mathcal{D}(g)$ an irrep of $G$, and similarly for the annihilation operators
\begin{equation}
\label{eq:psi_transformation}
\mathcal{U}_x(g)\psi_x \mathcal{U}^\dagger_x(g)=\mathcal{D}(g)^*\psi_x=\mathcal{D}(g^{-1})\psi_x\;,
\end{equation}
where we have assumed the irrep $\mathcal{D}(g)$ to be unitary.

Due to the anti-commuting nature of fermionic operators, the only way of constructing $\mathcal{U}_x(g)$ is the following:
\begin{equation}
    \mathcal{U}_x(g) = e^{i\alpha(g,x) \psi^\dagger_x \psi_x + i \beta(g, x)}\;,
\end{equation}
where $\alpha(g,x),\beta(g,x)$ are two constants.
Moreover, since we are considering $\mathbb{Z}_N$, we have that $\mathcal{D}(g^r) = \mathcal{D}(g)^r$ for integer $r$ and we can then focus our attention to the generator $g$ of the group.

Let us explicitly compute what is written in Eq.~\eqref{eq:psi_dag_transformation}:
\begin{equation}
    \mathcal{U}_x(g)\psi_x^\dagger \mathcal{U}^\dagger_x(g) = e^{i\alpha(g,x) \hat{n}_x} \psi_x^\dagger e^{-i\alpha(g,x) \hat{n}_x}
\end{equation}
where we denoted $\hat{n}_x = \psi^\dagger_x \psi_x$ the number operator. By applying this operator to a state $\ket{n}$, with $n=0,1$, we have:
\begin{equation}
    e^{i\alpha(g,x) \hat{n}_x} \psi_x^\dagger e^{-i\alpha(g,x) \hat{n}_x} \ket{n} = e^{i\alpha(g,x)} \psi_x^\dagger \ket{n}
    \label{eq:relation_for_Dg}
\end{equation}
It then follows that $\mathcal{D}(g) = e^{i\alpha(g,x)}$ and, since $\mathcal{D}(g)^N = 1$, we have $\alpha(g,x) = 2\pi r_g/N$ for some integer $r_g$. If we now take the generator $a$ of the group, so that for every element $g\in G$ we have $g=a^r$ for some integer $r=0,\dots,N-1$, we have that $\mathcal{D}(a)=\omega=\exp(i2\pi/N)$ so that $\alpha(g=a^r,x) = 2\pi r/N$.

In order to fix $\beta(g,x)$, we require the vacuum state $\ket{\Omega}$ on each site to transform trivially under the group. Thus, even sites should transform trivially if there are no particles (represented by $\ket{0}=\ket{\Omega_e}$), which means that we want:
\begin{equation}
    \mathcal{U}_x(g) \ket{0} = \ket{0}\;.
\end{equation}
To have this, we need $\beta(g,x) = 0$ for even sites. For odd sites instead, we want the state with no anti-particles (which is encoded by $\ket{1}=\ket{\Omega_o}$) to transform trivially
\begin{equation}
    \mathcal{U}_x(g) \ket{1} = \ket{1}\;,
    \label{eq:transformation_odd_sites}
\end{equation}
and we need $\beta(g,x)=-\alpha(g,x)$. To sum up, we can write for the generator $a$
\begin{equation}
    \mathcal{U}_x(a) = e^{i\frac{2\pi}{N}\mathcal{Q}_x}\;,
\end{equation}
where the staggered charge is
\begin{equation}
\label{eq:stag_charge_op}
    \mathcal{Q}_x = \psi_x^\dagger \psi_x - \frac{1}{2}\left(1 - (-1)^{|x|}\right)\;.
\end{equation}
For other group elements we have $\mathcal{U}_x(a^r)=\mathcal{U}_x(a)^r$.

In this framework, the fermionic Hamiltonian as defined in Eq.~\eqref{eq:free_fermionic_hamiltonian} is invariant under a global transformation, given by the application of $\mathcal{U}_x(g)$ on every site $x$. In order to extend this symmetry to a local symmetry, we need to define the so called parallel transporter. First, consider the hopping term of the free Hamiltonian. This transforms as
\begin{equation}
\begin{split}
    \mathcal{U}_x(g_x)\psi^\dagger_x \mathcal{U}^\dagger_x(g_x) &\mathcal{U}_y(g_y) \psi_y \mathcal{U}^\dagger_y(g_y) \\
    &= \mathcal{D}(g_x) \mathcal{D}(g_y^{-1}) \psi^\dagger_x \psi_y \\
    &= \mathcal{D}(g_x g_y^{-1}) \psi^\dagger_x \psi_y\;. \\
\end{split}
\end{equation}
We then introduce the operator $Q_{x,y}$, that is commonly called parallel transporter, where the group $G$ can act either from the left or the right as follows
\begin{equation}
\begin{split}
\mathcal{U}^L_{x,y}(g) Q_{x,y} {\mathcal{U}^{L\dagger}_{x,y}}(g) &= \mathcal{D}(g^{-1}) Q_{x,y}\\
\mathcal{U}^R_{x,y}(g) Q_{x,y} {\mathcal{U}^{R\dagger}_{x,y}}(g) &=  Q_{x,y}\mathcal{D}(g)\;,
\end{split}
\end{equation}
where $\mathcal{D}$ is the same irrep used for the fermionic matter.
If we then act on $Q_{x,y}$ with $g$ on the left and $h$ on the right we find it transforms as
\begin{equation}
    \mathcal{U}^L_{x,y}(g)\mathcal{U}^R_{x,y}(h) Q_{xy} {\mathcal{U}^{R\dagger}_{x,y}}(h){\mathcal{U}^{L\dagger}_{x,y}}(g) = \mathcal{D}(g^{-1} h) Q_{x,y}\;.
\end{equation}
When the group $G$ is Abelian, as in our case, the left and right transformations are adjoint of each other $\mathcal{U}^L_{x,y}(g)={\mathcal{U}^{R\dagger}_{x,y}}(g)=\mathcal{U}^L_{x,y}(g^{-1})$. To simplify the notation, we will then consider a single unitary $\mathcal{U}_{x,y}(g)\equiv\mathcal{U}^L_{x,y}(g)$ such that $\mathcal{U}^R_{x,y}(g)=\mathcal{U}^\dagger_{x,y}(g)$.

On a cubic lattice in $D$ spatial dimensions, every site $x$ has $2D$ nearest neighbor and thus the same number of operators $Q_{p,q}$ where wither $x=p$ or $x=q$. We can then think of the parallel transporters $Q_{x,y}$ as being associated to the links in the lattice. If we denote by $e_k$ the $D$ dimensional standard basis with $k=1,\dots,D$ we can identify two sets of links $(p,q)$ attached to a site $x$: incoming links have $q=x$ and $p=x-e_k$ for all $D$ choices of $k$, while outgoing links have $p=x$ and $q=x+e_k$.

With these definitions in place, we are now ready to define a unitary representation of the group $G$ associated to the site $x$ as
\begin{equation}
\label{eq:gen_G_law}
\mathcal{G}_x(g) =  \mathcal{U}_x(g) \prod_k \mathcal{U}_{x,x+e_k}(g)\mathcal{U}^\dagger_{x-e_k,x}(g)\;,
\end{equation}
this is the local gauge transformation we are looking for. It is in fact easy to see that, if we then modify the hopping term by adding the parallel transporter as
\begin{equation}
    \psi^\dagger_x \psi_y \;\rightarrow\; \psi^\dagger_x Q_{xy} \psi_y\;,
\end{equation}
the resulting Hamiltonian will now commute with $\mathcal{G}_x(g)$ for any site $x$ and group element $g$ individually. The full Hilbert space is then decomposed into separate sectors corresponding to different eigenvalues of these operators. The sector with no static charges is the one where these transformations act trivially so that a physical state $\ket{phys}$ will satisfy
\begin{equation}
\label{eq:gen_glaw}
\mathcal{G}_x(g) \ket{phys}=\ket{phys}\quad \forall x,\;\forall g\in G\;.
\end{equation}
This constraint is what is usually referred to as Gauss' law. Of course if we impose that
\begin{equation}
\mathcal{G}_x(a) \ket{phys}=\ket{phys}\quad \forall x\;,
\end{equation}
for the generator $a$ of $G$, than Eq.~\eqref{eq:gen_glaw} is automatically satisfied for every element of the group.

The parallel transporters $Q_{x,y}$ and the transformations $\mathcal{U}_{x,y}(g)$ act on states defined on the links of the lattice: the gauge fields. These also carry a representation of the group $G$. One possible choice is to take the eigenbasis of the parallel transporter $Q_{x,y}$ on the link such that
\begin{equation}
Q_{x,y}\ket{g} = \mathcal{D}(g)\ket{g}\;,
\end{equation}
where we used a direct representation of the group elements $\ket{g}$ for $g\in G$ using a link space of size $|G|$. In this basis, we can find how the link transformations act by considering the following
\begin{equation}
\begin{split}
Q_{x,y}\mathcal{U}_{x,y}(g)\ket{h}&=\mathcal{D}(g^{-1})\mathcal{U}_{x,y}(g)Q_{x,y}\ket{h}\\
&=\mathcal{D}(gh)\mathcal{U}_{x,y}(g)\ket{h}\;.
\end{split}
\end{equation}
But then, apart from a global phase we can choose to be 1, we have
\begin{equation}
\mathcal{U}_{x,y}(g)\ket{h} = \ket{gh}\;.
\end{equation}
In this basis, often called 'magnetic', the parallel transporters are diagonal and the transformations are not. We can also go to the 'electric' basis spanned by $\ket{q}$ for $q=0,\dots,|G|-1$ using a group Fourier transform
\begin{equation}
\langle g=a^r\vert q\rangle = \frac{1}{\sqrt{|G|}}\mathcal{D}(g)^q=\frac{1}{\sqrt{|G|}}\mathcal{D}(a)^{rq}\;,
\end{equation}
where the right hand side is in terms of the generator $a$. In the transformed basis we have instead that
\begin{equation}
\begin{split}
\mathcal{U}_{x,y}(g)\ket{r} &= \mathcal{D}(g)^r\ket{r}\\
Q_{x,y}\ket{r}&=\ket{r+1\mod |G|}\;.
\end{split}
\end{equation}
This is the basis we employ in the rest of this work.
%since $Q_{x,y}$ will cancel the $D(g_xg^{-1}_y)$ factor. \luca{$Q_{x,y}$ should have label $g$. Links? Gauss' Law?}

Finally, in order to make the gauge fields dynamic, we can add additional gauge invariant terms in the Hamiltonian expressed in terms of the $Q_{x,y}$ and $\mathcal{U}_{x,y}(a)$ operators. Since the latter commute with all the Gauss' law operators $\mathcal{G}_x(g)$ we can in principle include any functional of these provided the resulting operator is Hermitian. With the $Q_{x,y}$ operators instead, we can create gauge invariant operators by taking products along a closed path on the lattice. The smallest path being a plaquette.

\subsection{Encoding of gauge operators}

As a starting point to present the mapping we employ for a $\mathbb{Z}_N$ gauge theory coupled to fermions we consider first the following 1-dimensional model~\cite{Horn79}
\begin{equation}
\begin{split}
\label{eq:1dham}
    H &= m \sum_l (-1)^l \psi_l^{\dagger}\psi_l \\
    &+ \epsilon \sum_{l} \left( \psi_l^{\dagger} Q_{l,l+1} \psi_{l+1} + \psi_{l+1}^{\dagger} Q_{l+1,l} \psi_{l} \right) \\
    &-\lambda \sum_{l} \left( P_{l,l+1} + P_{l,l+1}^\dag \right)\;,
\end{split}
\end{equation}
where $l$ runs over the $L$ sites of the lattice, $Q_{l,l'}$ is the parallel transporter between site $l$ and site $l'$ (as defined in the previous section) and we denoted by $P$ the operator acting on the link between $l$ and $l'$
\begin{equation}
P_{l,l'} = \mathcal{U}_{l,l'}(a)\;,
\end{equation}
where $a$ is the generator of $\mathbb{Z}_N$. The position-dependent phases $s_{xy},s_x$ due to the fermionic staggering, in one dimensions, have been chosen to be $s_{xy}=1$, and $s_{x}=(-1)^x$. Following the discussion above, the $P$ and $Q$ operators satisfy
\begin{equation}
    P^N = Q^N = 1 \quad\quad\quad P Q P^\dag = \omega Q \;,
\end{equation}
and in the electric basis for the link act as
\begin{equation}
\begin{split}
    Q \ket{m} &= \ket{m+1 \mod N} \\
    P\ket{m} &= \omega^{m}\ket{m} \;,
\end{split}
\end{equation}
with $\omega=\exp(i2\pi/N)$.

It is then natural to encode the link variables using $N$ level qudits, one for each link, and express the operators $P$ and $Q$ in terms of the generalized Pauli matrices (see Eq.~\eqref{eq:generalized_pauli} in the appendix) 
\begin{equation}
\label{eq:link_map}
Q=\Xd,\quad P=\Zd\quad\rightarrow\quad \Zd\Xd\Zd^\dagger=\omega \Xd\;.
\end{equation}

The same construction carries over directly to arbitrary spatial dimensions $D$ since the gauge fields are defined on the links only.

\subsection{Fermionic encoding}
\label{sec:fermionic_encoding}

As for fermions, we will use for simplicity the Jordan-Wigner (JW) transformation~\cite{jw} to encode staggered fermions on the lattice. In more than 1 spatial dimension this results in a non-local Hamiltonian, other fermionic encoding could be used in principle to avoid this problem but we do not explore the issue further here. In order to simplify the error-correction scheme we will discuss later, we find it convenient to modify the standard JW mapping to use $N$-level qudits. Before describing this generalization we start by briefly reviewing the basic JW mapping applied to qubits. First, we define the creation and annihilation operators $a^\dag$ and $a$ respectively for every site $l$ of the lattice
\begin{equation}
\begin{split}
\label{eq:cran_ops}
    a_{l}^\dag &= \ketbra{1}{0}_{l} = \frac{1}{2} \left( 1 - Z_{l} \right) X_{l} \\
    a_{l} &=\ketbra{0}{1}_{l} = \frac{1}{2} \left( 1 + Z_{l} \right) X_{l} \;.
\end{split}
\end{equation}
For $D=1$ the label $l$ is directly the site label while for the general case we can pick a fixed ordering of sites on the lattice and use $l$ as the index for the corresponding site. These operators have the right anti-commutation relation on the same site but commute instead on different sites. The idea behind the JW mapping is that we can recover the correct fermionic anti-commutation relations by defining
\begin{equation}
    \psi_{l}^\dag = \left( \prod_{i=0}^{l-1} -Z_{i} \right)a_l^\dag \quad\text{and}\quad  \psi_l = \left( \prod_{i=0}^{l-1} -Z_{i} \right)a_l \;,
\end{equation}
where the product over $i$ runs over the sites preceding $l$ in the fixed ordering we chose. In this way we recover the right commutation relation between fermionic operators:
\begin{equation}
\begin{split}
\label{eq:anticomm}
    \{ \psi_{l}, \psi_{m} \} &= \{ \psi_{l}^\dag, \psi_{m}^\dag \} = 0 \\
    \{ \psi_{l}, \psi_{m}^\dag \} &= \delta_{l, m} \;.
\end{split}
\end{equation}

Now, we want to extend this to the generalized Pauli matrices acting on $N$-level systems. As we mentioned in the previous sections, when talking about error correction for qudits we ask $N$ to be a prime number. The generalized Pauli matrices are a complete basis of $N\times N$ matrices, in particular, considering only the generalized Pauli $\Zd$ matrix and its powers, it is a complete basis for diagonal matrices. These matrices satisfy the following orthonormality condition
\begin{equation}
    \frac{1}{N}Tr \left( \ketbra{i}{i}\Zd^{-j} \right) = \frac{\omega^{-ij}}{N}\;.
\end{equation}
Given this relation, we can write the projector on the $i$-th state as follows
\begin{equation}
    \ketbra{i}{i} = \frac{1}{N}\sum_{j=0}^{N-1} \omega^{-ij} \Zd^j\label{eq:projDecomp}\;.
\end{equation}

Under the assumption that we wish to use a single error correcting code to describe the Fermions and the link variables, it is necessary to embed the fermionic space as a qubit space within an $N$-dimensional qudit.
Following this strategy, we can use Eq.~\eqref{eq:projDecomp} and Eq.~\eqref{eq:pauliCommutation} to write the fermionic creation and annihilation operators as:
\begin{equation}
\begin{split}
    b^\dag &= \ketbra{1}{0} = \Xd \ketbra{0}{0} = \frac{1}{N}\sum_{j=0}^{N-1}\omega^{-j} \Zd^j \Xd \\
    b &= \ketbra{0}{1} = \ketbra{0}{0}\Xd^\dag = \frac{1}{N}\sum_{j=0}^{N-1}\Zd^j \Xd^\dag \\
\end{split}  
\end{equation}
Now, as in the qubit case, we need to add a string of operators that anti-commute with those $b^\dag$ and $b$, to restore the right commutation relation also on different sites. An operator that has the right requirements is the qubit ($2$-level) $Z$ Pauli matrix, that can be embedded in the $N$-level qudit operator space as $\widetilde Z = Z \oplus 0_{N-2}$, where $0_{N-2}$ is a $N-2$ by $N-2$ matrix of zeros. To write this operator in terms of generalized $N$-level Pauli matrices, we can use once again the projector notation:
\begin{equation}
\label{eq:ztilde}
    \widetilde Z = \ketbra{0}{0} - \ketbra{1}{1} = \frac{1}{N}\sum_{j=0}^{N-1} (1-\omega^{-j})\Zd^j
\end{equation}
By definition, this operator anti-commutes with both $b^\dag$ and $b$:
\begin{equation}
    b^\dag \widetilde Z + \widetilde Z b^\dag = 0 \quad\text{and}\quad
    b \widetilde Z + \widetilde Z b = 0 \\
\end{equation}
In exact analogy to the qubit case.

Denoting with $\widetilde Z_{l}$ the $\widetilde Z$ operator applied on the $l$-th site, we can define the fermionic creation and annihilation operators embedded in a $N$-level qudit space through a Jordan-Wigner like fermion to qudit mapping:
\begin{equation}
    \widetilde{\psi}_{l}^\dag = \left( \prod_{i=0}^{l-1} -\widetilde Z_{i} \right)b_l^\dag \quad\text{and}\quad
    \widetilde{\psi}_l = \left( \prod_{i=0}^{l-1} -\widetilde Z_{i} \right)b_l \;,
    \label{eq:generalized_jordan_wigner}
\end{equation}
which satisfy the anti-commutation relations in Eq.~\eqref{eq:anticomm}.

The operators defined above in Eq.~\eqref{eq:generalized_jordan_wigner} work correctly in the physical subspace spanned by states $\ket{0}$ and $\ket{1}$ and are otherwise trivial on all the other computational states. An alternative mapping can be found by requiring the fermionic operator to act correctly only when applied to physical states and leave the behavior otherwise arbitrary on the rest. For instance, consider the operators
\begin{equation}
\begin{split}
c^\dagger &= \left(1-\Zd\right)\frac{\Xd}{(1-\omega)^2} \left(\Zd-\omega\right)\\
c &= \left(\Zd^\dagger-\omega^{-1}\right)\frac{\Xd^\dagger}{(1-\omega^{-1})^2} \left(1-\Zd^\dagger \right)\;,
\end{split}
\end{equation}
which are one the adjoint of each other. For qubits $N=2$ and $\omega=\omega^{-1}=-1$ so these operators exactly match the standard ones from Eq.~\eqref{eq:cran_ops}. For generic $N$ instead these still have the correct behavior on physical states
\begin{equation}
\begin{split}
c^\dagger\ket{0} &= \ket{1},\quad c^\dagger\ket{1}=0,\quad c\ket{1} = \ket{0},\quad c\ket{0}=0\;,
\end{split}
\end{equation}
and do not couple these states to $\ket{q}$ for $q>1$. In order to use them to express fermionic operators we can either use the $\widetilde Z$ operators defined above or alternatively introduce
\begin{equation}
    \widetilde{\mathbf{Z}}=\frac{2}{1-\omega}\left(\frac{1+\omega}{2}-\Zd\right)\;,
\end{equation}
which anticommutes with both $c$ and $c^\dagger$ in the subspace spanned by $\ket{0}$ and $\ket{1}$. With this new operator, we avoid the need of an explicit summation over the $N$ levels as in Eq.~\eqref{eq:ztilde}, while still correctly giving $\widetilde{\mathbf{Z}}^\dagger\widetilde{\mathbf{Z}}=\mathbb{1}$ on the physical subspace.
We can then obtain the following set of fermionic operators
\begin{equation}
    \widetilde{\Psi}_{l}^\dag = \left( \prod_{i=0}^{l-1} \widetilde {\mathbf{Z}}^\dagger_{i} \right)c_l^\dag \quad\text{and}\quad
    \widetilde{\Psi}_l = \left( \prod_{i=0}^{l-1} \widetilde{\mathbf{Z}}_{i} \right)c_l \;.
    \label{eq:generalized_jordan_wigner_second}
\end{equation}

The main issue with these implementations of the phases operators is that they are not unitary. This means that in more than one spatial dimension the cost of block encoding will in general scale exponentially in the system size. One way to avoid this issue is to introduce instead a manifestly unitary operator like
\begin{equation}
\label{eq:zeta_JW}
\zeta=\exp\left(i\pi\frac{\mathcal{Z}-\mathcal{Z}^\dagger}{\omega-\omega^{-1}}\right)\;.
\end{equation}
This is well defined for $N>2$, anticommutes with $c$ and $c^\dagger$ and can be implemented as a block encoding deterministically using oblivious amplitude amplification~\cite{Berry_2014}.

\subsection{Hamiltonian in terms of generalized Pauli matrices}

Having set up the encoding of both gauge and matter degrees of freedom on a collection of $N$ level systems, we can now express the Hamiltonian in Eq.~\eqref{eq:1dham} using the generalized Pauli matrices. We start with the mapping of fermions in terms of $\widetilde{\psi}_{l}^\dag$ and $\widetilde{\psi}_{l}$ from Eq.~\eqref{eq:generalized_jordan_wigner}. The on site density operator reads
\begin{equation}
\widetilde{\psi}_{l}^\dagger\widetilde{\psi}_{l}=b_l^\dagger b_l = \ketbra{1}{1}_l = \frac{1}{N} \sum_{j=0}^{N-1} \omega^{-j}\Zd^j_l\;.
\end{equation}
On the other hand, the fermionic part of the hopping term reads
\begin{equation}
\begin{split}
\label{eq:hopping_first}
\widetilde{\psi}_{l}^\dagger\widetilde{\psi}_{l+1}&=-b_l^\dagger\widetilde{Z}_l b_{l+1}=-b_l^\dagger b_{l+1}\\
&=-\frac{1}{N^2}\sum_{j,k=0}^{N-1} \omega^{-j} \Zd_l^j \Zd_{l+1}^k \Xd_l \Xd_{l+1}^\dag\;,
\end{split}
\end{equation}
where in the first line we used that $b_l^\dagger$ act non trivially only on the state $\ket{0}$. Finally, the gauge links are naturally encoded into qubits using Eq.~\eqref{eq:link_map}. The $\mathbb{Z}_N$ Hamiltonian in 1 dimension can then be written as
\begin{equation}
\begin{split}
    &H = \frac{m}{N} \sum_l (-1)^l \sum_{j=0}^{N-1} \omega^{-j}\Zd^j_l \\
    &- \frac{\epsilon}{N^2} \sum_l \sum_{j,k=0}^{N-1} \left(  \omega^{-j}\Zd_l^j \Zd_{l+1}^k \Xd_l \Xd_{l,l+1} \Xd_{l+1}^\dag +\text{h.c.}\right)\\
    &-\lambda \sum_l \left(\Zd_{l,l+1} + \Zd_{l,l+1}^\dag\right)\;.
\end{split}
\end{equation}
The projection operators in the hopping term lead to the appearance of $O(LN^2)$ terms in this expansion of the Hamiltonian. As discussed in the previous section, this can be mitigated by using instead the fermionic encoding from Eq.~\eqref{eq:generalized_jordan_wigner_second}. In this case the on site occupation reads
\begin{equation}
\widetilde{\Psi}_{l}^\dag\widetilde{\Psi}_{l} =\frac{(1-\Zd_l)(\Zd_l-\omega^2)(\Zd^\dagger_l-\omega^{-2})(1-\Zd^\dagger_l)}{(1-\omega)^2(1-\omega^{-1})^2}\;.
\end{equation}
If we restrict the action of this operator on the physical subspace on site $l$ we see that this is equivalent to
\begin{equation}
\widetilde{\Psi}_{l}^\dag\widetilde{\Psi}_{l} \;\to\; \frac{(1-\Zd_l)(1-\Zd^\dagger_l)}{(1-\omega)(1-\omega^{-1})}=\frac{(1-\Zd_l)(1-\Zd^\dagger_l)}{4\sin^2(\pi/N)}\;.
\end{equation}
For the hopping term instead we have
\begin{equation}
\begin{split}
\widetilde{\Psi}_{l}^\dagger\widetilde{\Psi}_{l+1}&=c_l^\dagger\widetilde{\mathbf{Z}}_l c_{l+1}=-c_l^\dagger c_{l+1}\;,
\end{split}
\end{equation}
and the resulting expression in terms of generalized Pauli matrices contains now only $16$ terms instead of $N^2$ as in Eq.~\eqref{eq:hopping_first}. This form will then be convenient for $N>3$.

\subsection{Gauss's law embedded in $N$-level systems}
The key feature that we wish to study is whether the local symmetry corresponding to Gauss' law still allows us to build a qudit bit flip code if we take $N>2$.
We now will move closer to this goal by formulating the Gauss' law in terms of the generalized Pauli matrices.

In the pure gauge case, ie. in the absence of metter on sites, the Gauss law from Eq.~\eqref{eq:gen_G_law} can be written as
\begin{equation}
\begin{split}
\mathcal{G}_x(a) &=  \prod_k \mathcal{U}_{x,x+e_k}(a)\mathcal{U}^\dagger_{x-e_k,x}(a)\\
&=  \prod_k \Zd_{x,x+e_k}\Zd^\dagger_{x-e_k,x}\;,
\end{split}
\end{equation}
where $a$ is the generator of the $\mathbb{Z}_N$ group and in the second line we have expressed the link transformations in terms of generalized Pauli matrices.%If we identify the group generated by $\mathcal{G}_x(a)$ on all the $L$ sites with the stabilizer group, we directly obtain a $[dL,(d-1)L,3]_N$ classical bit-flip code in $d$ spatial dimensions.

The addition of fermionic matter on the sites creates a complication due to the fact that the site transformation
\begin{equation}
\mathcal{U}_x(a) = \exp\left(i\frac{2\pi}{N}\mathcal{Q}_x\right)\;,
\end{equation}
where $\mathcal{Q}_x$ is the staggered charge operator from Eq.~\eqref{eq:stag_charge_op}, cannot be expressed directly as products of generalized Pauli matrices. For instance, on the even sites we have
\begin{equation}
\mathcal{U}_x(a) = \omega^{\psi_x^\dagger\psi_x}=\begin{pmatrix}
    1 & 0 \\ 0 & \omega \\
    \end{pmatrix} \oplus M\;,
\end{equation}
where $M$ is an $N-2$ by $N-2$ diagonal matrix whose entries depend on how the encoded number operator $n_x=\psi_x^\dagger\psi_x$ is made to act on the additional qudit states. For instance, using the $\widetilde{\psi}^\dagger_x$ and $\widetilde{\psi}_x$ operators from Eq.~\eqref{eq:generalized_jordan_wigner} $M$ is the identity matrix. The diagonal elements would instead be different complex phases if we used the encoding operators $\widetilde{\Psi}^\dagger_x$ and $\widetilde{\Psi}_x$ from Eq.~\eqref{eq:generalized_jordan_wigner_second} instead.

The issue is inherently connected with the fact we are embedding the 2 dimensional space describing fermions as a subspace of the $N$ dimensional space of a qudit.

In order to overcome this difficulty, we consider instead a modified transformation acting on even sites as
\begin{equation}
\overline{\mathcal{U}}_x(a) = \omega^{\mathcal{N}_x}=\Zd_x\;,
\end{equation}
where we introduced a number operator acting as
\begin{equation}
\mathcal{N}_x \ket{n} = n \ket{n}\;,
\end{equation}
on the site qudit. These operators still satisfy the correct conjugation relations Eq.~\eqref{eq:psi_dag_transformation} and Eq.~\eqref{eq:psi_transformation} but with the qudit encoding of the fermionic operators instead.

The general transformation unitary will then take the following form
\begin{equation}
\overline{\mathcal{U}}_x(a)=\exp\left(i\frac{2\pi}{N}\overline{\mathcal{Q}}_x\right)\;,
\end{equation}
with a modified charge operator
\begin{equation}
\label{eq:qbar_def}
\overline{\mathcal{Q}}_x=\mathcal{N}_x-\frac{1}{2}\left[1-(-1)^{|x|}\right]\;.
\end{equation}
This operator matches exactly the staggered charge $\mathcal{Q}_x$ when acting on the physical subspace spanned by states $\ket{0}$ and $\ket{1}$ on the sites.

We can now consider the following modified Gauss laws
\begin{equation}
\begin{split}
\overline{\mathcal{G}}_x(a) &= \overline{\mathcal{U}}_x(a) \prod_k \mathcal{U}_{x,x+e_k}(a)\mathcal{U}^\dagger_{x-e_k,x}(a)\\
&= \omega^{-p_x}\Zd_x \prod_k \Zd_{x,x+e_k}\Zd^\dagger_{x-e_k,x}\;,
\label{eq:modified_gauss_law}
\end{split}
\end{equation}
where for convenience we have defined
\begin{equation}
    p_x := \frac{1}{2}\left[ 1-(-1)^{|x|} \right]
\end{equation}
to account for the staggering of charges.

Every gauge invariant state in the theory is stabilized by these modified Gauss law, in the sense that every physical state is a $+1$ eigenvalue of $\overline{\mathcal{G}}_x(a)$ for every $x$. However, because we embedded the $2$-dimensional fermionic space on each site into a $N$-dimensional Hilbert space, we also have a remnant gauge symmetry constraining the physical state on each site to be only $\ket{0}$ or $\ket{1}$.

We can express such symmetry as
\begin{equation}
\begin{split}
\label{eq:gxpi_def}
    \mathcal{G}_x^\pi &= \begin{pmatrix}
        1 & 0 \\
        0 & 1 \\
    \end{pmatrix} \oplus M \\
\end{split}
\end{equation}
for any $(N-2)\times(N-2)$ matrix $M$ with no $+1$ eigenvalue. A possible option is to identify this operator with the generator $\mathcal{G}_x^\pi(a)$ of a residual $\mathbb{Z}_2$ symmetry. This can be done by choosing
\begin{equation}
\begin{split}
\label{eq:gxpiofa}
\mathcal{G}_x^\pi(a)&=-\exp\left(i\pi(\ketbra{0}{0} + \ketbra{1}{1})\right)\\
&=-\exp\left(i\frac{\pi}{N}\sum_{j=0}^{N-1}(1+\omega^{-j})\Zd^j_x\right)\;,
%&=-\exp\left(i\frac{\pi}{2N}\sum_{j=0}^{N-1}\left((1+\omega^{-j})\Zd^j_x+(1+\omega^{j})\Zd^{-j}_x\right)\right)\\
\end{split}
\end{equation}
where we used Eq.~\eqref{eq:projDecomp} to express the projectors.

Of course, this operator and $\overline{\mathcal{G}}_x(a)$ commute, which means that we can define the common physical space to both Gauss laws ($\overline{\mathcal{G}}_x(a)$ and $\mathcal{G}_x^\pi(a)$), and it easy to see that this will be equivalent to the physical space of the original Gauss law $\mathcal{G}_x(a)$.

As a final comment we want to mention that the modified Gauss laws $\overline{\mathcal{G}}_x(a)$ are exactly the local gauge transformations for the situation where matter on the sites is represented as an $N$-level boson with associated charge operator $\overline{\mathcal{Q}}_x$ from Eq.~\eqref{eq:qbar_def} above. This is often referred as "Higgs matter" in the literature~\cite{PhysRevD.19.3682,PhysRevResearch.3.013133} and, as we can see from the construction presented above, the gauge redundancy of this theory can be completely captured by the qudit stabilizer formalism. It is the addition to two-level fermions on the site that gives rise to the appearence of the residual $\mathbb{Z}_2$ Gauss' Law  $\mathcal{G}_x^\pi$ from Eq.~\eqref{eq:gxpi_def}.

\section{Gauss's law error correcting code}
\label{section:4}

The modified Gauss's law operator in Eq.~\eqref{eq:modified_gauss_law} has been written as a product of generalized Pauli $\Zd$ matrices, and so it is in the generalized Pauli group. All of the Gauss's law operators $\overline{\mathcal{G}}_l(a)$ commute with each other (as expected for an Abelian theory). It therefore follows that a stabilizer group $S$ can be defined, through the group generated by $\overline{\mathcal{G}}_l(a)$ for every site $l$ in the system. Considering $D$ spatial dimensions and $M$ sites, we have $DM$ links, $M$ sites, and $M$ stabilizers ($1$ per site). This means that the code will have parameters $[[(D+1)M, DM, d]]_N$ (where we use the notation $[[n,k,d]]_p$, where $n$ is the number of physical qudits, $k$ the number of logical qudits, $d$ is the distance and $p$ the number of levels of the qudit, as defined in Section~\ref{section:2}).

Since the Hamiltonian commutes with every $\overline{\mathcal{G}}_l(a)$ by construction, and since the logical operations of a code are a complete basis for operators that commute with $S$, it is possible to write the Hamiltonian in terms of the logical operations of the code defined by $S$.

The question now is what is the distance $d$, and to find out this, by definition of $d$ we have to look at the logical operations. Starting with the logical $\Xd$ operations, one can see that one possible choice of operators is $\bar \Xd_l = \Xd_l \Xd_{l,l+1} \Xd^\dag_{l+1}$, which is exactly the off-diagonal term in the Hamiltonian. This operator commutes with every Gauss's law operator. Since we have one such operator for every link, there are $DM$ logical $\Xd$. As for the logical $\Zd$ operators, we can choose $\bar \Zd_l = \Zd_{l,l+1}$. This operator commutes with every Gauss's law, commutes with $\bar \Xd_m$ for every $m\ne l$, and has the right commutation relation on the same logical qudit $\bar \Xd_l \bar \Zd_l = \omega^{-1} \bar \Zd_l \bar \Xd_l$.

This is not the only possible choice of logical operations, but we consider this choice because it leads to logical operators that have the minimum possible weight. Indeed, from the definition of those logical operations, one can see that the distance of the quantum error correcting code is $d=1$ (since the $Z$ logical operations have weight $1$), and the code is able to correct, in general, $0$ errors. For clarity, here the distance is computed as the minimum between the distance $d_x$ and $d_z$ (see Section~\ref{section:2}). However, we can distinguish between the distance concerning $\Xd$ errors $d_x$, and the one concerning $\Zd$ errors $d_z$, since asymmetric distances are very useful (and concatenation is always possible to increase one of the two distances). Again, we have that $d_z = 1$ and the code generated by $G_l$ cannot correct or detect any $\Zd$ errors. However, as $d_x = 3$, the code is able to detect and correct every single-qubit $\Xd$ error. This generalizes the result seen in the qubit case \cite{spagnoli2024faulttolerantsimulationlatticegauge} to the qudit case.

Specifically, the stabilizer group generated by $\overline{\mathcal{G}}_l$ generates a classical bit-flip code $[(D+1)M, DM, 3]_N$, and if we concatenate the code with a qudit phase flip code then we 
obtain a full quantum error correcting code with parameters $[[3(D+1)M,DM,3]]$.  This is a substantial improvement over a na\"ive generalization of the Shor code for qudits by a factor of $3$.

Another reason for which we chose those operators as logical operators is that they are a natural choice of operators to write the Hamiltonian $H$:
\begin{equation}
\begin{split}
    \overline{\mathcal{G}}_l(a) &= \omega^{-p_l} \Zd_{l-1,l} \Zd_l \Zd_{l,l+1}^\dag \\
    \bar \Xd_l &= \Xd_l \Xd_{l,l+1} \Xd^\dag_{l+1} \\
    \bar \Zd_l &= \Zd_{l,l+1} \\
    \Zd_l &= \omega^{p_l} G_l \bar \Zd_{l-1}^\dag \bar \Zd_l \\
\end{split}
\end{equation}
Now, we can substitute those operators in the Hamiltonian, and neglecting the Gauss's law since it is the logical identity we get the Hamiltonian in terms of logical operators:
\begin{equation}
\begin{split}
    H &= \frac{m}{N} \sum_l (-1)^l \sum_{j=0}^{N-1} \omega^{-j} \omega^{jp_l} \bar \Zd_{l-1}^{-j} \bar \Zd_l^{j} \\
    &- \frac{\epsilon}{N^2} \sum_l \sum_{j,k=0}^{N-1} \left( \omega^{-j} \omega^{jp_l+kp_{l+1}} \bar \Zd_{l-1}^{-j} \bar \Zd_{l}^{j-k} \bar \Zd_{l+1}^{k} \bar \Xd_l \right. \\
    &\quad\quad\quad\quad\quad+ \left. \omega^{-k} \omega^{jp_l+kp_{l+1}} \bar \Zd_{l-1}^{-j} \bar \Zd_{l}^{j-k} \bar \Zd_{l+1}^{k} \bar \Xd_l^\dag \right) \\
    &-\lambda \sum_l \bar \Zd_{l} + \bar \Zd_{l}^\dag
\end{split}
\end{equation}

The original Hamiltonian had, as degrees of freedom, $M$ sites and $DM$ links. However, due to the Gauge symmetry, not all of them were independent. Now, we've been able to express the Hamiltonian in terms of only logical (ie. gauge invariant) degrees of freedom, which now are instead all independent. What we achieved is equivalent to have integrated out the Gauge constraint, leaving in the Hamiltonian only the independent (and new) degrees of freedom which are ``collective", in the sense that they cannot be always mapped $1$ to $1$ with the old links and sites, but they could encompass multiple links and sites.

As mentioned in the previous section, there is a remnant gauge symmetry, forcing fermions to occupy only the first two levels of every qudit representing a site of the lattice. This symmetry can be expressed via the operator $\mathcal{G}^\pi_l(a)$, which is not in the generalized Pauli group, and therefore cannot be used as additional stabilizer. However, the Hamiltonian commutes with $\mathcal{G}^\pi_l(a)$ for every $l$, which means that the time evolution will not mix the first two levels on a site with the other levels. This then imply that only a bit-flip error can bring the evolution from the physical space in the unphysical space corresponding of a site occupying a higher level. Due to the fact that the Gauss' law $\overline{\mathcal{G}}_l(a)$ generates a classical bit-flip code, if such a bit-flip error occurs, it can be corrected with the bit-flip code generated by $\overline{\mathcal{G}}_l(a)$. We conclude that the additional symmetry generated by $\mathcal{G}^\pi_l(a)$ is not needed to do a fault-tolerant simulation of the system we are considering, assuming that the initial state is initialized to be a $+1$ eigenvector of $\mathcal{G}^\pi_l(a)$ for every $l$.

\section{Universal gate set}
\label{section:universal_gate_set}

In this section, we will show how one could implement a universal set of gates on the gauge covariant error-correcting code we introduced above employing a generalziation of the strategy proposed in Ref.~\cite{spagnoli2024faulttolerantsimulationlatticegauge} for the $N=2$ case. In particular, we will consider the Clifford+T gate set as defined in Section~\ref{section:2}, with the Clifford group generated by the $\SUMgate$, single-qubit $QFT$ and $\Sd$ gates.

The gauge covariant error-correcting code generated by the $\mathbb{Z}_N$ Gauss Law operators has an asymmetry between $X$ and $Z$ stabilizers (indeed, there are no $X$ stabilizers). This means that the code is not self-dual, and due to this fact, we cannot have a transversal logical Hadamard gate (or in our qudit case, a transversal logical QFT). However, we can inject the QFT by exploiting an external qudit. Consider the following circuit (where we denoted with L the outcome of the measurement):
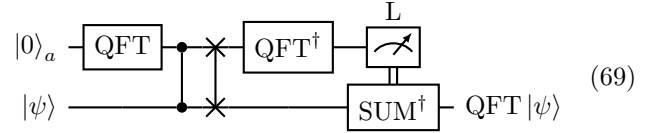
\begin{equation}
\begin{quantikz}[row sep={0.8cm,between origins},column sep={0.2cm}]
    \lstick{$\ket{0}_a$} & \gate{\QFT} & \ctrl{1} & \swap{1} & \gate{\QFT^\dag} & \meter{L}\vcw{1} \\
    \lstick{$\ket{\psi}$} & \qw & \ctrl{0} & \swap{0} & \qw & \gate{\SUMgate^\dag} & \qw \rstick{$\QFT \ket{\psi}$} \\
\end{quantikz}
\end{equation}
In Appendix~\ref{App:state_injection} we prove that this circuit actually works when considering both $\ket{0}_a$ and $\ket{\psi}$ as single qudits. To translate this into our error-correcting code, we can consider the ancilla qudit to be encoded in some CSS qudit error-correcting code on which it is possible to implement the QFT operator, and that has transversal operations with our Gauss's law error-correcting code. For example, consider the bit-flip repetition code, which corrects single-qudit $X$ errors. It is a classical code, as the Gauss' Law code, and when concatenating the Gauss' Law code with another code to make it quantum, it is enough to use the same concatenation also in the ancilla. For example, consider the ancilla in the bit-flip code, and the concatenation of both the system and the ancilla with a 3-qudit phase-flip code. This would result in the ancilla being encoded in the 9-qudit Shor's code, that has transversal CNOT and CZ with the Gauss' Law code, but on which it may be difficult to apply the QFT operation. If instead we consider the $7$-qudit code, that has transversal $\overline{X}$ with the Gauss' law code, but not transversal $\overline{Z}$, it would be easy to apply the QFT, but one should find the right concatenation to make $\overline{Z}$ transversal as well. In Appendix~\ref{sec:qudit_phase_flip_code} we provide the example of a qudit phase-flip code, which could be used (in the right basis) as code in the ancilla to make the CNOTs transversal. Then, since our code encodes multiple logical qudits and we cannot identify a logical qudit with a specific set of physical qudits, we can consider the logical operations. If we want to apply the $\QFT$ on the $k$-th logical qudit, we simply have to substitute every $\Xd$ and $\Zd$ operation with the logical $\bar \Xd_k$ and $\bar \Zd_k$ on that logical qudit (also in the controlled operations). For the argument of before, all controlled operations and $\text{SWAP}$ gate will remain transversal between the $3$-qudit repetition code and the Gauss's law code.

It is important to notice that, in order to perform the $\text{SWAP}$ gate, we need either the $\widetilde{CX}$ gate, or the $K$ gate (both defined in Appendix~\ref{sec:gates_and_measurements}). Since the $\widetilde{CX}$ gate is defined through the $\QFT$ which is the gate we are trying to apply, using this gate to perform the $\text{SWAP}$ would not then provide a usable implementation. Thus, let us consider the definition of the $\text{SWAP}$ gate based on the $\SUMgate$ gate. This implementation normally requires the $K$ gate, defined as $K\ket{j} = \ket{-j}$, which is in general difficult to realize. However, the swap gate is followed by a $\QFT^\dagger$, and we can exploit the following relation:
\begin{equation}
    K\QFT \ket{j} = \QFT^\dag \ket{j}
\end{equation}
Thus, the circuit becomes:
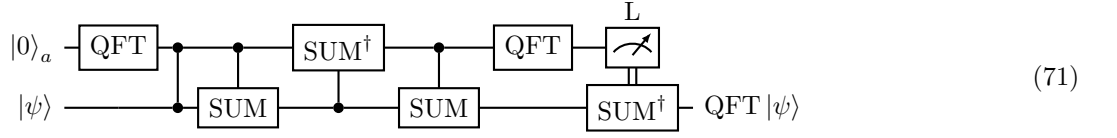
\begin{widetext}
\begin{equation}
\begin{quantikz}[row sep={0.8cm,between origins},column sep={0.2cm}]
    \lstick{$\ket{0}_a$} & \gate{\QFT} & \ctrl{1} & \ctrl{1} & \gate{\SUMgate^\dag} & \ctrl{1} & \gate{\QFT} & \meter{L}\vcw{1} \\
    \lstick{$\ket{\psi}$} & \qw & \ctrl{0} & \gate{\SUMgate} & \ctrl{-1} & \gate{\SUMgate} & \qw & \gate{\SUMgate^\dag} & \qw \rstick{$\QFT \ket{\psi}$} \\
\end{quantikz}
\label{eq:H_injection}
\end{equation}
\end{widetext}
where the first $C\Zd$ gate can be also written as a $\SUMgate$ by using the $\QFT$.

The same logic applies for the non-Clifford $T$ gate. Following Ref.~\cite{PhysRevA.98.052108}, for every diagonal unitary $U$, the circuit shown below can be used for qubits:
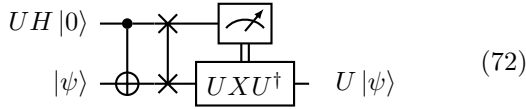
\begin{equation}
\begin{quantikz}[row sep={0.8cm,between origins},column sep={0.2cm}]
    \lstick{$UH\ket{0}$} & \ctrl{1} & \swap{} & \meter{}\vcw{1} \\
    \lstick{$\ket{\psi}$} & \targ{} & \swap{-1} & \gate{UXU^\dagger} & \qw & \rstick{$U\ket{\psi}$}  \\
\end{quantikz}
\label{circuit:U_qubit_injection}
\end{equation}
It can be generalized to the qudit case as
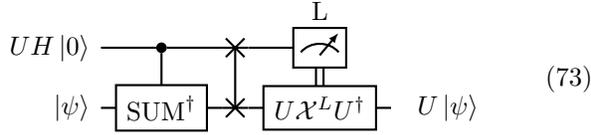
\begin{equation}
\begin{quantikz}[row sep={0.8cm,between origins},column sep={0.2cm}]
    \lstick{$UH\ket{0}$} & \ctrl{1} & \swap{} & \meter{L}\vcw{1} \\
    \lstick{$\ket{\psi}$} & \gate{\SUMgate^\dagger} & \swap{-1} & \gate{U\Xd^LU^\dagger} & \qw & \rstick{$U\ket{\psi}$}  \\
\end{quantikz}
\label{circuit:U_qudit_injection}
\end{equation}
The calculation for both the qubit and qudit circuits is in Appendix \ref{App:state_injection}.

This circuit is easy to apply in our case, since the definition of $\Td$ gate that we introduced in Eq.~\eqref{eq:definition_T} (see also~\cite{jain2020normalformsinglequditcliffordt}) is such that $\Td \Xd \Td^\dagger$ is Clifford (we prove this in Appendix \ref{App:cliffordST}). In this case, the $\text{SWAP}$ gate can be easily realized with the $\SUMgate$ since the single qubit unitary $K$, which is in general non-trivial, would happen just before the measurement and due to the fact that it simply maps one state to another, it can be computed in post-processing, without actually applying it to the ancilla. This means that the circuit to inject the diagonal gate $U$ becomes:
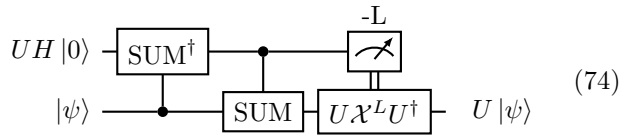
\begin{equation}
\begin{quantikz}[row sep={0.8cm,between origins},column sep={0.2cm}]
    \lstick{$UH\ket{0}$} & \gate{\SUMgate^\dag} & \ctrl{1} & \meter{-L}\vcw{1} \\
    \lstick{$\ket{\psi}$} & \ctrl{-1} & \gate{\SUMgate} & \gate{U\Xd^LU^\dagger} & \qw & \rstick{$U\ket{\psi}$}  \\
\end{quantikz}
\end{equation}

This means that both the $\QFT$ and the $\Td$ gates can be applied on our Gauss's law error correcting code by using state injection, where the ancilla qudit can be, for example, in the $7$-qudit error correcting code, while on our system only transversal Clifford gates are required. Since the CNOT (or the $\SUMgate$) is transversal by definition of CSS code, the result is that we can in this way apply a universal set of gates on the Gauge covariant code in a fault-tolerant way.

\section{Logical duality}
\label{section:5}

As a direct application of the formalism developed in the previous sections, we now show  that the Hamiltonian can be written in terms of bosonic operators only. Since we start from the logical Hamiltonian, and we arrive to a dual theory, we will call this ``logical duality", which results in a theory that lacks all fermionic degrees of freedom. This simplifies the description of the logical Hamiltonian, showing that encoding the system with the Gauge-covariant error correcting code is equivalent to integrating out fermions from the system. Integrating out fermions was already possible for theories where the Gauge group had $\mathbb{Z}_2$ as a normal subgroup (otherwise, the symmetry group had to be extended to have this property) \cite{PhysRevB.98.075119} or paying the price of extending the range of the interaction \cite{PhysRevD.99.114511}. Here, we show how to integrate out the matter degrees of freedom in theories with a $\mathbb{Z}_p$ Gauge group, where $p$ is a prime different to $2$. Our current construction still relies on the Jordan-Wigner mapping to map fermions into hard-core bosons leading to non-localities in more than one spatial dimension.

In this section we consider in detail the case of $D=1$ spatial dimensions and comment on the extension to larger $D$ in the next subsection.

Let us define the following bosonic operators acting on the $l$-th logical qudit:
\begin{equation}
\begin{split}
    \phi_l^\dag &= \sum_{n=0}^{N-2} \sqrt{n+1} \ket{n+1}\bra{n} \\
    \phi_l &= \sum_{n=0}^{N-2} \sqrt{n+1} \ket{n}\bra{n+1} \\
\end{split}
\end{equation}
These operators are the standard bosonic creation and annihilation operators, with the addition of a cutoff that sets the boundary condition $\phi^\dag \ket{N-1} = 0$. The infinite-dimensional bosonic Hilbert space with all standard properties and relations is restored by taking the limit of $N\rightarrow \infty$. It is easy to prove all usual relations, starting with the number operator
\begin{equation}
    n_l = \phi_l^\dag \phi_l = \sum_{n=0}^{N-1} n \ket{n}\bra{n}
\end{equation}
and the bosonic commutation relations
\begin{equation}
\begin{split}
    [\phi_l, \phi_m] &= [\phi_l^\dag, \phi_m^\dag] = 0 \\
    [\phi_l, \phi_m^\dag] &= \left( \mathbb{1} - N\ket{N-1}\bra{N-1} \right)\delta_{l,m} \\
\end{split}
\end{equation}
Note that, due to the finite value of $N$, we have an additional boundary term in the commutators.

With these definitions, we can write the generalized logical Pauli matrices as follows:
\begin{equation}
\begin{split}
\label{eq:bos_to_Pauli}
    \bar \Zd_l &= e^{i\frac{2\pi}{N} n_l} \quad\quad
    \bar \Xd_l = \QFT\ \bar \Zd_l\ \QFT^\dag \\
\end{split}
\end{equation}
where the $\QFT$ operator is meant to be on the logical degree of freedom. If we call $\rho_l$ the Fourier transform of the on-site number operator%and $\pi^\dag$ the conjugate momenta, defined as
\begin{equation}
\rho_l =  \QFT\  n_l \ \QFT^\dagger\;,
\end{equation}
we can write the logical $\Xd$ operation as the exponential:
\begin{equation}
    \bar \Xd_l = e^{i\frac{2\pi}{N} \rho_l}\;.%\pi_l^\dag \pi_l} \;.
\end{equation}

This expression makes it apparent that the number operator in Fourier space $\rho_l$ is proportional to the generator of the incremementer in the bosonic Fock space.

Now, we are ready to write the Hamiltonian in terms of these new bosonic degrees of freedom. Since we already showed how to write $\bar \Xd_l$ and $\bar \Zd_l$ in terms of $\phi_l$ and $\pi_l$, one could simply substitute those in the Hamiltonian and get a bosonic Hamiltonian. However, we can perform several simplifications, and to show that, let us start from the mass term in terms of logical operations. If we substitute those logical operations with their definition in terms of bosonic operators we get the following expression for the fermionic density
\begin{equation}
\begin{split}
\label{eq:fmass_boson}
\widetilde{\psi}_l^{\dagger}\widetilde{\psi}_l&=\frac{1}{N}\sum_{j=0}^{N-1} \omega^{-j} \omega^{jp_l} \bar \Zd_{l-1}^{-j} \bar \Zd_l^{j}\\
&= \frac{1}{N} \sum_{j=0}^{N-1} e^{i\frac{2\pi}{N}j (n_l-n_{l-1}+p_l-1)}\\
&=\delta_N(n_l - n_{l-1} + p_l - 1) \;,
\end{split}
\end{equation}
where $\delta_N$ is the Kronecker delta modulo N. An equivalent expression can be found using the alternative fermionic encoding from Eq.~\eqref{eq:generalized_jordan_wigner_second} leading to
\begin{equation}
\widetilde{\Psi}_l^{\dagger}\widetilde{\Psi}_l = \frac{\sin^2((n_l+p_l-n_{l-1})\pi/N)}{\sin^2(\pi/N)}\;.
\end{equation}
The two separate definitions match in the physical subspace of fermions on sites while only the expression in Eq.~\eqref{eq:fmass_boson} is zero outside of this subspace. For later convenience we introduce 
\begin{equation}
\label{eq:pidelta}
\pi_l = \delta_N(n_l - n_{l-1} + p_l - 1)\;,
\end{equation}
which is manifestly a projection operator.
The staggered mass term in the Hamiltonian takes then the form
\begin{equation}
\begin{split}
m \sum_l (-1)^l \widetilde{\psi}_l^{\dagger}\widetilde{\psi}_l&=m \sum_l (-1)^l \pi_l\;.
\end{split}
\end{equation}
The electric term in the Hamiltonian can be written instead simply as
\begin{equation}
-\lambda \sum_l \left(\bar{\Zd}_l+\bar{\Zd}^\dagger_l\right)=-2\lambda\sum_l \cos\left(\frac{2\pi}{N}n_l\right)\\
\end{equation}
taking the form of an on-site interaction for the bosons.

Using the same procedure one can also show that the hopping term, in the physical subspace, takes the form
\begin{equation}
\widetilde{\psi}_l^\dagger Q_{l,l+1}\widetilde{\psi}_{l+1}=\pi_le^{i\frac{2\pi}{N} \rho_l}\pi_{l+1}\;,
\end{equation}
while it's complex conjugate gives
\begin{equation}
\widetilde{\psi}_{l+1}^\dagger Q^\dagger_{l,l+1}\widetilde{\psi}_{l}=\pi_{l+1}e^{-i\frac{2\pi}{N} \rho_l}\pi_{l}\;
\end{equation}
The full hopping term can now be written as follows
\begin{equation}
\begin{split}
H_h &= \sum_l \left(\widetilde{\psi}_l^\dagger Q_{l,l+1}\widetilde{\psi}_{l+1}+\widetilde{\psi}_{l+1}^\dagger Q^\dagger_{l,l+1}\widetilde{\psi}_{l}\right)\\
&=\sum_l \left(\pi_le^{i\frac{2\pi}{N} \rho_l}\pi_{l+1}+\pi_{l+1}e^{-i\frac{2\pi}{N} \rho_l}\pi_{l}\right)
\end{split}
\end{equation}

The full Hamiltonian for a $\mathbb{Z}_N$ theory with staggered fermions in $D=1$ spatial dimensions can the be expressed as
\begin{equation}
\begin{split}
\label{eq:full1dham}    
H=&-2\lambda\sum_l \cos\left(\frac{2\pi}{N}n_l\right)\\
&+m \sum_l (-1)^l \pi_l\\
&-\epsilon\sum_l \left(\pi_le^{i\frac{2\pi}{N} \rho_l}\pi_{l+1}+\pi_{l+1}e^{-i\frac{2\pi}{N} \rho_l}\pi_{l}\right)
\end{split}
\end{equation}

It is important to notice that the degrees of freedom are defined in terms of bosonic operators, which are in general very hard to implement do to the square roots in the coefficients. However, those operators appear in the Hamiltonian only in pairs, either as the number operator $n_l$, or its conjugate operator $\rho_l$, and those are much easier to implement.

The calculation for the $2$-dimensional lattice is equivalent, by defining $\phi_{\mathbf{l}, \mu}$ the bosonic degrees of freedom associated to the link that starts form site $\mathbf{l}$ in the direction $\mu$. In this setting, the Kronecker delta $\pi_l$ introduced by the local symmetry and defined in Eq.~\eqref{eq:pidelta} above need to be generalized to
\begin{equation}
\pi_{\mathbf{l}}=\delta_N( n_{\mathbf{l},x} + n_{\mathbf{l},y} -n_{\mathbf{l}-x,x} - n_{\mathbf{l}-y,y} +p_{\mathbf{l}} - 1)\;,
\end{equation}
in order to account for the 4 links connected to the $\mathbf{l}$ matter site. The purely gauge part of the Hamiltonian has a similar electric term as before, and now also a plaquette term
\begin{equation}
\begin{split}
H_{gauge}=&-2\lambda_E\sum_{\mathbf{l}, \mu} \cos\left( \frac{2\pi}{N} n_{\mathbf{l}, \mu} \right)\\
-2\lambda_P\sum_{\mathbf{l}} &\cos\left( \frac{2\pi}{N} \left(\rho_{\mathbf{l}, x}+\rho_{\mathbf{n+x}, y}-\rho_{\mathbf{l}, y}-\rho_{\mathbf{n+y}, x}\right) \right)\;.
\end{split}
\end{equation}
The fermionic mass term is completely equivalent to the one dimensional case apart from the convention for staggered fermions
\begin{equation}
H_M=m \sum_{\mathbf{l}} (-1)^{l_x+l_y}\pi_{\mathbf{l}}\;.
\end{equation}
Finally, the hopping term in the $x$ direction is completely analogous to the one dimensional case, apart from the different definition of the delta operators $\pi_{\mathbf{l}}$, while on the $y$ direction we need to take into account the Jordan-Wigner strings. The result reads
\begin{equation}
\begin{split}    
H_{hop} = &-\epsilon \sum_{\mathbf{l}}\left(\pi_{\mathbf{l}}e^{i\frac{2\pi}{N}\rho_{\mathbf{l}}}\pi_{\mathbf{l}+x}+\pi_{\mathbf{l}+x}e^{-i\frac{2\pi}{N}\rho_{\mathbf{l}}}\pi_{\mathbf{l}}\right)\\
& -\epsilon \sum_{\mathbf{l}}P_{\mathbf{l}}\left(\pi_{\mathbf{l}}e^{i\frac{2\pi}{N}\rho_{\mathbf{l}}}\pi_{\mathbf{l}+y}+\pi_{\mathbf{l}+y}e^{-i\frac{2\pi}{N}\rho_{\mathbf{l}}}\pi_{\mathbf{l}}\right)\;.
\end{split}
\end{equation}
where the phase operator $P_{\mathbf{l}}$ is
\begin{equation}
P_{\mathbf{l}} = (-1)^{\sum_{\mathbf{k}=\mathbf{l}}^{\mathbf{l}+y}(1-\pi_{\mathbf{k}})}
\end{equation}
A full derivation of these results for the $2+1$ dimensional case can be found in Appendix~\ref{app:2D_bosonization}.

The above Hamiltonians have the great virtue of being more compact and being defined on a vector space where the $\mathbb{Z}_N$ Gauss law is trivial at every site. This is true both in the bosonic representation used in this section as well as using directly the logical generalized Pauli operators. The version best used for simulations of the theory might then depend on the specific platform employed: e.g. on qudit devices with accurate and fast generalized Pauli operations the latter version might be preferable. In contrast with the $N=2$ case, already explored in Refs.~\cite{Rajput2023,spagnoli2024faulttolerantsimulationlatticegauge}, we are not able here to completely remove the local redundancy from the theory: when we represent fermionic matter using only the first two levels of a $N$-level qudit we implicitly introduce an additional $\mathbb{Z}_2$ symmetry for the theory at every site. As shown above, this symmetry is generated by the operator $\mathcal{G}_x^\pi(a)$ from Eq.~\eqref{eq:gxpiofa} which in one dimension can be rewritten as
\begin{equation}
\begin{split}
\mathcal{G}_l^\pi(a) &= -\exp\left(i\frac{\pi}{N}\sum_{j=0}^{N-1}(1+\omega^{-j})\omega^{jp_l}\bar \Zd^{-j}_{l-1}\bar \Zd^{j}_{l}\right)\\
&= -\exp\left(i\pi\left(\delta_N(n_l - n_{l-1} + p_l - 1)\right.\right.\\
&\quad\quad\quad\quad\quad\;\;\left.+\left.\delta_N(n_l - n_{l-1} + p_l)\right)\right)\;,
\end{split}
\end{equation}
and analogously for higher spatial dimension. When the system is initialized in the correct subspace the dynamics is maintained inside of it. In alternative, for instance when considering the preparation of low-energy thermal states, one can add an energy penalty to the Hamiltonian using for instance
\begin{equation}
H_{gauss} = \Lambda \left(1-\mathcal{G}_l^\pi(a)\right)\;,
\end{equation}
or another alternative would be to use
\begin{equation}
\begin{split}
H'_{gauss} = \Lambda &\left(1-\delta_N(n_l - n_{l-1} + p_l - 1)\right)\\
&\times\left(1-\delta_N(n_l - n_{l-1} + p_l )\right)\;,
\end{split}
\end{equation}
both with a positive coupling constant $\Lambda$.
With such a penalty terms the unphysical site vector space will be gapped in energy from the correct one.

\section{Conclusions}

In this work, we have extended the paradigm of gauge-covariant error correction from qubits to qudits, providing a unifying framework for the fault-tolerant digital simulation of Abelian lattice gauge theories on cyclic groups of prime dimension. By embedding Gauss's law directly into the stabilizer structure, we constructed codes that not only preserve gauge invariance but also reduce the number of physical degrees of freedom, mirroring and generalizing the result achieved for $\mathbb{Z}_2$ in the qubit case \cite{Rajput2023,spagnoli2024faulttolerantsimulationlatticegauge}. The identification of logical operators enabled us to rewrite the encoded Hamiltonian entirely in terms of independent logical degrees of freedom, opening the door to more resource-efficient simulations.

A key outcome of this work is the identification of a logical duality: starting from the logical Hamiltonian—namely, the Hamiltonian expressed in terms of logical degrees of freedom—we map it to two distinct bosonic models, preserving the locality of the physical interaction. Both bosonic models, as well as the logical Hamiltonian itself, demonstrate that matter located on sites can be integrated out without extending the range of the interaction or introducing redundant degrees of freedom. This represents a concrete advantage for both one- and two-dimensional systems. Although the two models originate from the same logical Hamiltonian, they involve a trade-off. The second model is in general cheaper to implement, since its decomposition in terms of generalized Pauli is much simpler. However, this model does not completely remove the gauge symmetry: a residual symmetry remains, and must be enforced either in terms of stabilizer operators, or by adding a penalty term in the Hamiltonian. By using error correction, it is possible to say that if errors are only single-qubit, and we are able to correct all single-qubit errors (as in the case of the Gauss law error-correcting code we introduced), then this residual symmetry will never be violated, and thus it can be neglected.

This paves the way to a more general framework to find dualities between gauge theories, exploiting the formalism or quantum error correction. Both gauge theories and error-correcting codes are based on the definition of constraints. By connecting multiple gauge theories to the same error correcting code either proves their equivalence, or allow us to write the residual symmetries to map one theory into another. 

Finally, we showed how a universal fault-tolerant gate set can be realized by combining transversal Clifford operations with state-injection protocols between compatible qudit codes, thereby ensuring computational universality without sacrificing error protection.    

%A key outcome is the logical duality that maps the encoded lattice gauge theory onto two different bosonic models while preserving the locality of the interaction. This bosonization demonstrates that matter on sites can be integrated out without extending the interaction range or introducing redundant degrees of freedom, an advance that directly benefits both one- and two-dimensional systems. This paves the way to a more general framework to find dualities between gauge theories, by exploiting the formalism or quantum error correction. At the same time, we showed how a universal fault-tolerant gate set can be realized by combining transversal Clifford operations with state-injection protocols between compatible qudit codes, thereby ensuring computational universality without sacrificing error protection.

Although the physical interactions remain local in the bosonized Hamiltonian, the Jordan-Wigner Pauli strings remain nonlocal under the qudit mapping. This limitation can, in principle, be mitigated with alternative encodings. In particular, fermion-to-qudit mappings that exploit additional on-site levels can yield local fermionic representations in two spatial dimensions \cite{carobene2024localfermiontoquditmappings, vezvaee2024quantumsimulationfermihubbardmodel}. Exploring these approaches within our gauge-covariant framework is a promising direction for future work.

Taken together, our results establish a flexible pathway to symmetry-preserving, fault-tolerant simulations of lattice gauge theories on multilevel hardware. We expect the techniques introduced here to inform non-Abelian generalizations and hardware-tailored implementations, bringing scalable quantum simulations of high-energy and condensed-matter phenomena closer to practical realization.

\acknowledgments{
%We would like to thank ...

A.R. is funded by the European Union. Views and opinions expressed are however those of the author(s) only and do not necessarily reflect those of the European Union or the European Commission. Neither the European Union nor the granting authority can be held responsible for them. This project has received funding from the European Union’s Horizon Europe research and innovation programme under grant agreement No.\ 101080086 NeQST.

NW acknowledges the support from DOE, Office of Science, National Quantum Information Science Research Centers, Co-design Center for Quantum Advantage (C2QA) under Contract No.~DE-SC0012704 (Basic Energy Sciences, PNNL FWP 76274) and Pacific Northwest National Laboratory's Quantum Algorithms and Architecture for Domain Science (QuAADS) Laboratory Directed Research and Development (LDRD) Initiative. 

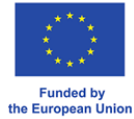
\begin{center}
    \includegraphics[width=0.1\textwidth]{EU.png}
\end{center}

}

\appendix
\onecolumngrid

\section{Clifford and non-Clifford gates}
\label{App:cliffordST}

In this section we will provide the proofs that the generalizations of the $S$ and $T$ gates as provided in the main text are Clifford and non-Clifford respectively.

Starting from the $\Sd$ gate, to be Clifford we have to prove that both $\Sd \Zd \Sd^\dag$ and $\Sd \Xd \Sd^\dag$ can be written as a product of Pauli matrices. The term with $\Zd$ is trivial since $[\Sd,\Zd]=0$, so $\Sd \Zd \Sd^\dag = \Zd$. The term with $\Xd$ instead is:
\begin{equation}
\begin{split}
    \Sd \Xd \Sd^\dag &= \left( \sum_{i=0}^{d-1} \omega^{i(i+1) [2^{-1}]_d} \ket{i}\bra{i} \right) \left( \sum_{j=0}^{d-1} \ket{j}\bra{j+1} \right) \left( \sum_{k=0}^{d-1} \omega^{-k(k+1) [2^{-1}]_d} \ket{k}\bra{k} \right) \\
    &= \left( \sum_{j=0}^{d-1} \omega^{j(j+1) [2^{-1}]_d} \ket{j}\bra{j+1} \right) \left( \sum_{k=0}^{d-1} \omega^{-k(k+1) [2^{-1}]_d} \ket{k}\bra{k} \right) \\
    &= \sum_{j=0}^{d-1} \omega^{j(j+1) [2^{-1}]_d} \omega^{-(j+1)(j+2) [2^{-1}]_d} \ket{j}\bra{j+1} \\
    &= \sum_{j=0}^{d-1} \omega^{-(2j-2) [2^{-1}]_d} \ket{j}\bra{j+1} \\
    &= \sum_{j=0}^{d-1} \omega^{-(j-1)} \ket{j}\bra{j+1} = \omega \Zd^\dag \Xd \\
\end{split}
\end{equation}

Considering instead the $\Td$ gate, the term with $\Zd$ is again trivial since $[\Td, \Zd]=0$, which means $\Td \Zd \Td^\dag = \Zd$. As for the $\Xd$ term we have:
\begin{equation}
\begin{split}
    \Td \Xd \Td^\dag &= \left( \sum_{i=0}^{d-1} \omega^{i^3 [6^{-1}]_d} \ket{i}\bra{i} \right) \left( \sum_{j=0}^{d-1} \ket{j}\bra{j+1} \right) \left( \sum_{k=0}^{d-1} \omega^{-k^3 [6^{-1}]_d} \ket{k}\bra{k} \right) \\
    &= \left( \sum_{j=0}^{d-1} \omega^{j^3 [6^{-1}]_d} \ket{j}\bra{j+1} \right) \left( \sum_{k=0}^{d-1} \omega^{-k^3 [6^{-1}]_d} \ket{k}\bra{k} \right) \\
    &= \sum_{j=0}^{d-1} \omega^{-(3j^2+3j+1) [6^{-1}]_d} \ket{j}\bra{j+1} \\
    &= \omega^{-[6^{-1}]_d} \sum_{j=0}^{d-1} \omega^{-j(j+1) [2^{-1}]_d} \ket{j}\bra{j+1} = \omega^{-[6^{-1}]_d} \Sd^\dag \Xd \\
\end{split}
\end{equation}
We use the fact that $[6^{-1}]_d \cdot 3 = [2^{-1}]_d$. For clarity, this can be shown as
\begin{equation}
1 = [6^{-1}]_d \cdot 6 
  = [6^{-1}]_d \cdot (3 \cdot 2) 
  = \bigl([6^{-1}]_d \cdot 3\bigr) \cdot 2,
\end{equation}
which implies $[6^{-1}]_d \cdot 3 = [2^{-1}]_d$. This proof requires \( [2^{-1}]_d \neq 0 \), which is trivial since we assumed \( d \) to be a prime greater than \(2\). It also requires \( [6^{-1}]_d \neq 0 \). 
This condition is less trivial, because it means that \( d \) must not divide \(6\), i.e., \( d \neq 2 \) and \( d \neq 3 \). In conclusion, the proof holds for every prime \( d \geq 5 \). In Appendix \ref{appendix:Td_qutrits} we show an example of why this does not work for $d=3$. As a conclusion, since $\Td \Xd \Td^\dag = \omega^{[6^{-1}]_d} \Sd^\dag \Xd$, it contains an $\Sd$ gate, which is not in the Pauli group, we can say that $\Td$ is not in the generalized Clifford group.

\section{State injection}
\label{App:state_injection}

In this appendix we will prove that the circuits provided in the main text do what they are claimed to do. In particular, we will show it for the circuits that perform the state injection of the QFT given in Eq.~\eqref{eq:H_injection}, the state injection of the general single-qubit diagonal unitary given in Eq.~\eqref{circuit:U_qubit_injection} and the corresponding qudit generalization of Eq.~\eqref{circuit:U_qudit_injection}.

\subsection{Quantum Fourier Transform}
Consider the circuit in Eq.~\eqref{eq:H_injection}, and let us recall some relations:
\begin{equation}
\begin{split}
    C\Zd &= \sum_{j=0}^{d-1} \ket{j}\bra{j} \otimes \Zd^j = \sum_{j=0}^{d-1} \ket{j}\bra{j} \otimes \sum_{k=0}^{d-1} \omega^{jk} \ket{k}\bra{k} \\
    \QFT &= \frac{1}{\sqrt{d}} \sum_{j=0}^{d-1} \sum_{k=0}^{d-1} \omega^{jk} \ket{j}\bra{k} \quad\quad\quad \ket{\psi} = \sum_{j=0}^{d-1} c_j \ket{j} \\
\end{split}
\end{equation}
Then, starting from the beginning of the circuit, and denoting with a subscript $a$ the ancilla qubit, we have:
\begin{equation}
\begin{split}
    \ket{\psi}\ket{0}_a =& \sum_{j=0}^{d-1} c_j \ket{j} \ket{0}_a \\
    \QFT \rightarrow & \sum_{j=0}^{d-1} c_j \ket{j} \frac{1}{\sqrt{d}} \sum_{k=0}^{d-1} \ket{k}_a \\
    CZ \rightarrow & \frac{1}{\sqrt{d}} \sum_{j,k=0}^{d-1} c_j \omega^{jk} \ket{j} \ket{k}_a \\
    \text{SWAP} \rightarrow & \frac{1}{\sqrt{d}} \sum_{j,k=0}^{d-1} c_j \omega^{jk} \ket{k}\ket{j}_a \\
    \QFT^\dag \rightarrow & \frac{1}{d} \sum_{j,k=0}^{d-1} c_j \omega^{jk}\ket{k}\sum_{l=0}^{d-1} \omega^{-jl} \ket{l}_a \\
    &= \frac{1}{d} \sum_{j=0}^{d-1} c_j \left( \sum_{k=0}^{d-1} \sum_{l=0}^{d-1} \omega^{j(k-l)} \ket{k} \ket{l}_a \right)
\end{split}
\end{equation}
Then, the ancilla qudit is measured and let us call $L$ the outcome of the measurement. Then, we classically control an operation $X^{-L}$ on the system qudit, so that:
\begin{equation}
\begin{split}
    \text{MEASUREMENT = L} \rightarrow & \frac{1}{\sqrt{d}} \sum_{j=0}^{d-1} c_j \left( \sum_{k=0}^{d-1} \omega^{j(k-L)} \ket{k} \right) \\
    X^{-L} \rightarrow & \frac{1}{\sqrt{d}} \sum_{j=0}^{d-1} c_j \left( \sum_{k=0}^{d-1} \omega^{j(k-L)} \ket{k-L} \right) \\
    &= \frac{1}{\sqrt{d}} \sum_{j=0}^{d-1} c_j \left( \sum_{k=0}^{d-1} \omega^{jk} \ket{k} \right) = \QFT \ket{\psi}
\end{split}
\end{equation}
In this way, we proved that the circuit in Eq.~\eqref{eq:H_injection} actually works.

\subsection{Qubit T-gate}
Consider the circuit in Eq.~\eqref{circuit:U_qubit_injection}. To prove that this circuit works, we can do the following computation, by writing $\ket{\psi} = a\ket{0} + b\ket{1}$ and $U\ket{j} = u_j\ket{j}$:
\begin{equation}
\begin{split}
    \ket{\psi} UH\ket{0}_a &= (a\ket{0}+b\ket{1})\frac{1}{\sqrt{2}}U(\ket{0}_a+\ket{1}_a) \\
    &= (a\ket{0}+b\ket{1})\frac{1}{\sqrt{2}}(u_0\ket{0}_a+u_1\ket{1}_a) \\
    CNOT &\rightarrow \frac{1}{\sqrt{2}} u_0 (a\ket{0}+b\ket{1}) \ket{0}_a + \frac{1}{\sqrt{2}} u_1 (a\ket{1}+b\ket{0}) \ket{1}_a \\
    &= \frac{1}{\sqrt{2}} (u_0  a\ket{0} \ket{0}_a + u_0 b\ket{1} \ket{0}_a + u_1  a\ket{1} \ket{1}_a + u_1 b\ket{0} \ket{1}_a) \\
    SWAP&\rightarrow \frac{1}{\sqrt{2}} (u_0  a\ket{0} \ket{0}_a + u_0 b\ket{0} \ket{1}_a + u_1  a\ket{1} \ket{1}_a + u_1 b\ket{1} \ket{0}_a) \\
    &= \frac{1}{\sqrt{2}} (u_0a\ket{0} +u_1b\ket{1}) \ket{0}_a + \frac{1}{\sqrt{2}} (u_1a\ket{1} +u_0b\ket{0}) \ket{1}_a \\
\end{split}
\end{equation}
If we now measure the ancilla qubit, we have $1/2$ probability to get zero, in which case the circuit would result in $U\ket{\psi}$, as we wanted, and $1/2$ probability to apply a different unitary.
In the case of the measurement outcome being $1$, we apply the $UXU^\dag$ unitary on the $\ket{\psi}$ register, and this will result, again, in the $U\ket{\psi}$ state on the main register, as we wanted.

\subsection{Qudit T-gate}
Consider the circuit in Eq.~\eqref{circuit:U_qudit_injection}. The previous computation can be generalized to the qudit case by writing $\ket{\psi} = \sum_j c_j\ket{j}$ and $U\ket{j} = u_j \ket{j}$. In this case, we have:
\begin{equation}
\begin{split}
    \ket{\psi}U\QFT \ket{0}_a &= \sum_{j=0}^{d-1} c_j \ket{j} U \frac{1}{\sqrt{d}} \sum_{k=0}^{d-1} \ket{k} \\
    &= \sum_{j=0}^{d-1} c_j \ket{j} \frac{1}{\sqrt{d}} \sum_{k=0}^{d-1} u_k \ket{k} \\
    SUM^\dag &\rightarrow \frac{1}{\sqrt{d}} \sum_{j=0}^{d-1} \sum_{k=0}^{d-1} c_j u_k \ket{j-k} \ket{k} \\
    SWAP &\rightarrow \frac{1}{\sqrt{d}} \sum_{j=0}^{d-1} \sum_{k=0}^{d-1} c_j u_k \ket{k} \ket{j-k} \\
    \text{MEASUREMENT = }L &\rightarrow \sum_{j=0}^{d-1} c_j u_{j-L} \ket{j-L} \\
    UX^LU^\dagger &\rightarrow UX^LU^\dagger \sum_{j=0}^{d-1} c_j u_{j-L} \ket{j-L} \\
    &= UX^L \sum_{j=0}^{d-1} c_j \ket{j-L} \\
    &= U \sum_{j=0}^{d-1} c_j \ket{j} \\
\end{split}
\end{equation}
In this way we proved that the equivalent qudit circuit is the following:
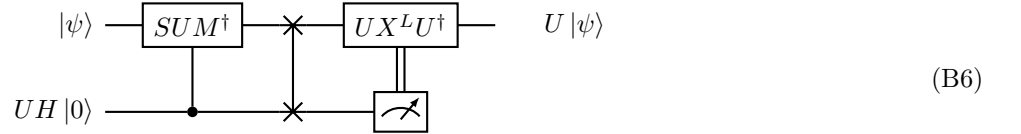
\begin{equation}
\begin{quantikz}
    \lstick{$\ket{\psi}$} & \gate{SUM^\dagger} & \swap{1} & \gate{UX^LU^\dagger} & \qw & \rstick{$U\ket{\psi}$}  \\
    \lstick{$UH\ket{0}$} & \ctrl{-1} & \swap{} & \meter{}\vcw{-1} \\
\end{quantikz}
\end{equation}
where the power of $X$, that we called $L$, is the outcome of the measurement.

\section{T gate for d=3}
\label{appendix:Td_qutrits}

Consider the circuit in Eq.~\eqref{circuit:U_qudit_injection}, used for injecting the $\Td$ gate. For that circuit to work, we need to be able to apply on the system the $U\Xd U^\dagger$ gate, and we can do that if it is Clifford. In appendix \ref{App:cliffordST}, we show that $\Td \Xd \Td^\dagger \propto \Sd^\dagger \Xd$, which means it is Clifford, but that proof only works for $d\ge 5$. In this section, we will show that such a simple choice of $\Td$ gate does not work for $d=3$. In particular, we will show that does not exists a diagonal unitary, with eigenvalues that are powers of $\omega_9 = (\omega_3)^{\frac{1}{3}}$, such that both the following holds:
\begin{itemize}
    \item $\Td$ is non-Clifford
    \item $\Td \Xd \Td^\dagger$ is Clifford
\end{itemize}
Remember that in the qutrit case we have:
\begin{equation}
    \Xd = \begin{pmatrix}
        0 & 1 & 0 \\
        0 & 0 & 1 \\
        1 & 0 & 0 \\
    \end{pmatrix}
    \quad\quad\quad
    \Sd = \begin{pmatrix}
        1 & 0 & 0 \\
        0 & \omega_3 & 0 \\
        0 & 0 & 1 \\
    \end{pmatrix}
\end{equation}
From here it is easy to see that all diagonal gates with eigenvalues that are powers of $\omega_3$ are Clifford. This means that we need another phase to define the non-Clifford $\Td$ gate, and that's why we chose a $\Td$ gate with eigenvalues that are powers of $\omega_9$. So let's assume:
\begin{equation}
    \Td = \begin{pmatrix}
        \omega_9^a & 0 & 0 \\
        0 & \omega_9^b & 0 \\
        0 & 0 & \omega_9^c \\
    \end{pmatrix}
\end{equation}
for some $a,b,c \in \mathbb{Z}$. Then, we can compute the unitary we are interested into as
\begin{equation}
\begin{split}
    \Td \Xd \Td^\dag &=\begin{pmatrix}
        \omega_9^a & 0 & 0 \\
        0 & \omega_9^b & 0 \\
        0 & 0 & \omega_9^c \\
    \end{pmatrix}
    \begin{pmatrix}
        0 & 1 & 0 \\
        0 & 0 & 1 \\
        1 & 0 & 0 \\
    \end{pmatrix}
    \begin{pmatrix}
        \omega_9^{-a} & 0 & 0 \\
        0 & \omega_9^{-b} & 0 \\
        0 & 0 & \omega_9^{-c} \\
    \end{pmatrix} \\
    &= \begin{pmatrix}
        0 & \omega_9^a & 0 \\
        0 & 0 & \omega_9^b \\
        \omega_9^c & 0 & 0 \\
    \end{pmatrix}
    \begin{pmatrix}
        \omega_9^{-a} & 0 & 0 \\
        0 & \omega_9^{-b} & 0 \\
        0 & 0 & \omega_9^{-c} \\
    \end{pmatrix} \\
    &= \begin{pmatrix}
        0 & \omega_9^{a-b} & 0 \\
        0 & 0 & \omega_9^{b-c} \\
        \omega_9^{c-a} & 0 & 0 \\
    \end{pmatrix} \\
    &= \begin{pmatrix}
        \omega_9^{a-b} & 0 & 0 \\
        0 & \omega_9^{b-c} & 0 \\
        0 & 0 & \omega_9^{c-a} \\
    \end{pmatrix} X
\end{split}
\end{equation}
This unitary, to be Clifford, has to be written in terms of only $\omega_3$. This means that every power of $\omega_9$ has to be a multiple of $3$. It follows that:
\begin{equation}
    \begin{pmatrix}
        a-b \\
        b-c \\
        c-a \\
    \end{pmatrix} = 
    \begin{pmatrix}
        0 \\
        0 \\
        0 \\
    \end{pmatrix} \mod\ 3
\end{equation}
Solving this system of equations, we find the condition $a=b=c$, which makes $T$ proportional to the identity, and in particular Clifford. This is the contradiction that proves the fact that with this choice of $\Td$ gate, it is impossible that $\Td \Xd \Td^\dag$ is Clifford. As a consequence, we cannot use the Circuit for state injection of the $\Td$ gate for $d=3$ (unless one finds a choice of $\Td$ gate with $d=3$ such that it is non-Clifford while $\Td \Xd \Td^\dag$ is Clifford).

\section{$2$-dimensional logical duality}
\label{app:2D_bosonization}

In this section we will show how the 2-dimensional Hamiltonian for a $\mathbb{Z}_N$ gauge theory with staggered fermions can be written in terms of only bosonic degrees of freedom. This is equivalent to finding a dual Hamiltonian which describes the same theory, but where there is no sign problem since there are no fermions. The organization is as follows: first, we will present the Hamiltonian, by writing it in terms of (generalized) Pauli matrices. Then, by defining the logical operations on the 2-dimensional Gauss's law code, we will write the same Hamiltonian in terms of only logical degrees of freedom. Finally, we will express the logical degrees of freedom in terms of bosonic operators.

Consider the two dimensional Hamiltonian as the sum of the mass, hopping, electric and plaquette terms:
\begin{equation}
    H = H_M + H_{hop} + H_E + H_P
\end{equation}

We will use a notation where $\mathbf{n} = \{ i,j \}$ enumerates every site in the lattice, while $\mu = \{ x,y \}$ denotes the possible directions in the lattice. Operators written with only one index, like $O_{\mathbf{n}}$, are meant to act on the matter site labeled by $\mathbf{n}$, while operators that have also the subscript of the direction $O_{\mathbf{n},\mu}$ are meant to act on the link which originate from the site $\mathbf{n}$ and is in the direction $\mu$. From this notation, it follows that the electric term is:
\begin{equation}
    H_E = -\lambda_E \sum_{\mathbf{n}, \mu} \left( P_{\mathbf{n},\mu} + P_{\mathbf{n},\mu}^\dag \right) = -\lambda_E \sum_{\mathbf{n}, \mu} \left( \Zd_{\mathbf{n}, \mu} + \Zd^\dag_{\mathbf{n}, \mu} \right)
\end{equation}
The plaquette term is straightforward as well:
\begin{equation}
    H_P = -\lambda_P \sum_{\mathbf{n}} \left( Q_{\mathbf{n},y} Q_{\mathbf{n+y},x} Q_{\mathbf{n+x},y}^\dag Q_{\mathbf{n},x}^\dag + \text{ h.c.}  \right) = -\lambda_P \sum_{\mathbf{n}} \left( \Xd_{\mathbf{n},y} \Xd_{\mathbf{n+y},x} \Xd_{\mathbf{n+x},y}^\dag \Xd_{\mathbf{n},x}^\dag + \text{ h.c.}  \right)
\end{equation}

In order to write the mass and hopping terms, we need to define the fermionic encoding in terms of Pauli matrices. As in the 1-dimensional case, we will use the Jordan-Wigner encoding defined in Section \ref{sec:fermionic_encoding}. In order to implement the fermionic mapping, we need to decide an ordering for sites. In the following calculations we will consider the ordering by rows, which means that if there are $N_s$ sites per row, the integer labeling the site with $\mathbf{n} = \{ n_x, n_y \}$ will be $f(\mathbf{n}) = l = n_x + n_yN_s$. Then, the fermionic operators can be written as:
\begin{equation}
\begin{split}
    \widetilde{\psi}_{\mathbf{n}}^\dag &= \left( \prod_{\mathbf{k}<\mathbf{n}} -\widetilde Z_{\mathbf{k}}  \right) b^\dag_{\mathbf{n}} \quad\quad
    \widetilde{\psi}_{\mathbf{n}} = \left( \prod_{\mathbf{k}<\mathbf{n}} -\widetilde Z_{\mathbf{k}}  \right) b_{\mathbf{n}} \;.
\end{split}
\end{equation}

We can now write the mass and hopping terms as
\begin{equation}
    H_M = m \sum_{\mathbf{n}} (-1)^{n_x+n_y} \psi^\dag_{\mathbf{n}} \psi_{\mathbf{n}} = \frac{m}{N} \sum_{\mathbf{n}} (-1)^{n_x+n_y} \sum_{j=0}^{N-1} \omega^{-j} \Zd_{\mathbf{n}}^j
\end{equation}
\begin{equation}
    H_{hop}^x = \epsilon \sum_{\mathbf{n}} \left( \psi^\dag_{\mathbf{n}} Q^\dag_{\mathbf{n}, x} \psi_{\mathbf{n}+x} + \text{ h.c.} \right) = -\frac{\epsilon}{N^2} \sum_{\mathbf{n}} \sum_{j,k=0}^{N-1} \left( \omega^{-j} Z^j_{\mathbf{n}} Z^k_{\mathbf{n}+x} \Xd_{\mathbf{n}} \Xd^\dag_{\mathbf{n}, x} \Xd^{\dagger}_{\mathbf{n}+x}  + \text{ h.c.} \right)
\end{equation}
\begin{equation}
    H_{hop}^y = \epsilon \sum_{\mathbf{n}} \left( \psi^\dag_{\mathbf{n}} Q^\dag_{\mathbf{n}, y} \psi_{\mathbf{n}+y} + \text{ h.c.} \right) = -\frac{\epsilon}{N^2} \sum_{\mathbf{n}} P_{\mathbf{n}} \sum_{j,k=0}^{N-1} \left( \omega^{-j} Z^j_{\mathbf{n}} Z^k_{\mathbf{n}+y} \Xd_{\mathbf{n}} \Xd^\dag_{\mathbf{n}, y} \Xd^{\dagger}_{\mathbf{n}+y}  + \text{ h.c.} \right)
\end{equation}
where $P_{\mathbf{n}}$ is the Pauli string coming from Jordan-Wigner for vertical links, defined as follows:
\begin{equation}
    P_{\mathbf{n}} = \prod_{\mathbf{k}=\mathbf{n}}^{\mathbf{n}+y} \left( -\widetilde Z_{\mathbf{k}} \right), \quad\quad\quad \widetilde Z_{\mathbf{k}} = \frac{1}{N} \sum_{j=0}^{N-1} \left( 1-\omega^{-j} \right) \Zd_{\mathbf{k}}^j
\end{equation}

Now, we have the Hamiltonian written as a function of Pauli matrices on sites and links. From here, we can define the logical operations for this system, by considering the Gauss's law code. If we have a total number of sites equal to $\mathcal{N}$, in a 2-dimensional spatial lattice we will have $2\mathcal{N}$ links. This means we need $3\mathcal{N}$ $N$-level qudits to encode the system, while the number of stabilizers (Gauss's law) will be one per site. We will then need $2\mathcal{N}$ logical $\bar X$ and $2\mathcal{N}$ logical $\bar Z$ operations. Since they are in the same number as the links, we can use the same notation we used to number links to number logical operations. One possible choice of logical operations, equivalent to the choice we did in the 1-dimensional case, is the following:
\begin{equation}
    \bar \Xd_{\mathbf{n}, \eta} = \Xd_{\mathbf{n}} \Xd_{\mathbf{n}, \eta}^\dag \Xd_{\mathbf{n} + \eta}^\dag \quad\quad
    \bar \Zd_{\mathbf{n}, \eta} = \Zd_{\mathbf{n}, \eta}
\end{equation}
In order to write the Hamiltonian in terms of those logical operations, we only need to write $\Zd_{\mathbf{n}}$ on sites in terms of logical $\bar \Zd_{\mathbf{n}}$. We can do it by using the definition of the Gauss's law in 2 spatial dimensions:
\begin{equation}
\begin{split}
    G_{\mathbf{n}} &= \omega^{-p_{\mathbf{n}}} \Zd_{\mathbf{n}} \Zd_{\mathbf{n}-x,x} \Zd_{\mathbf{n}-y,y} \Zd_{\mathbf{n},y}^\dag \Zd_{\mathbf{n},x}^\dag \\
    \Zd_{\mathbf{n}} &= \omega^{p_{\mathbf{n}}} G_{\mathbf{n}} \bar \Zd_{\mathbf{n}-x,x}^\dag \bar \Zd_{\mathbf{n}-y,y}^\dag \bar \Zd_{\mathbf{n},y} \bar \Zd_{\mathbf{n},x} \\
    p_{\mathbf{n}} &= \frac{1}{2}\left( 1 - (-1)^{n_x+n_y} \right) \\
\end{split}
\end{equation}
We can then write every term in the Hamiltonian in terms of those logical operations:
\begin{equation}
    H_M = \frac{m}{N} \sum_{\mathbf{n}} (-1)^{n_x+n_y} \sum_{j=0}^{N-1} \omega^{-j} \omega^{jp_{\mathbf{n}}} \bar \Zd_{\mathbf{n}-x,x}^{-j} \bar \Zd_{\mathbf{n}-y,y}^{-j} \bar \Zd_{\mathbf{n},y}^j \bar \Zd_{\mathbf{n},x}^j
\end{equation}
\begin{equation}
    H_{hop}^x = -\frac{\epsilon}{N^2} \sum_{\mathbf{n}} \sum_{j,k=0}^{N-1} \left( \omega^{-j} \omega^{jp_{\mathbf{n}}} \omega^{kp_{\mathbf{n}+x}} \bar \Zd_{\mathbf{n}-x,x}^{-j} \bar \Zd_{\mathbf{n}-y,y}^{-j} \bar \Zd_{\mathbf{n},y}^{j} \bar \Zd_{\mathbf{n},x}^{j-k} \bar \Zd_{\mathbf{n}+x-y,y}^{-k} \bar \Zd_{\mathbf{n}+x,y}^{k} \bar \Zd_{\mathbf{n}+x,x}^{k} \bar \Xd_{\mathbf{n}, x} + \text{ h.c.} \right)
\end{equation}
\begin{equation}
    H_{hop}^y = -\frac{\epsilon}{N^2} \sum_{\mathbf{n}} P_{\mathbf{n}} \sum_{j,k=0}^{N-1} \left( \omega^{-j} \omega^{jp_{\mathbf{n}}} \omega^{kp_{\mathbf{n}+y}} \bar \Zd_{\mathbf{n}-x,x}^{-j} \bar \Zd_{\mathbf{n}-y,y}^{-j} \bar \Zd_{\mathbf{n},x}^{j} \bar \Zd_{\mathbf{n},y}^{j-k} \bar \Zd_{\mathbf{n}+y-x,x}^{-k} \bar \Zd_{\mathbf{n}+y,x}^{k} \bar \Zd_{\mathbf{n}+y,y}^{k} \bar \Xd_{\mathbf{n}, y} + \text{ h.c.} \right)
\end{equation}
\begin{equation}
    P_{\mathbf{n}} = \prod_{\mathbf{k}=\mathbf{n}}^{\mathbf{n}+y} \left( -\frac{1}{N}\sum_{j=0}^{N-1} \left( 1-\omega^{-j} \right) \omega^{jp_{\mathbf{k}}} \bar \Zd^{-j}_{\mathbf{k}-x,x} \bar \Zd^{-j}_{\mathbf{k}-y,y} \bar \Zd^{j}_{\mathbf{k},y} \bar \Zd^{j}_{\mathbf{k},x} \right)
\end{equation}
\begin{equation}
    H_E = -\lambda_E \sum_{\mathbf{n}, \mu} \left( \bar \Zd_{\mathbf{n}, \mu} + \bar \Zd^\dag_{\mathbf{n}, \mu} \right)
\end{equation}
\begin{equation}
    H_P = -\lambda_P \sum_{\mathbf{n}} \left( \bar \Xd_{\mathbf{n},y}^\dag \bar \Xd_{\mathbf{n}+y,x}^\dag \bar \Xd_{\mathbf{n}+x,y} \bar \Xd_{\mathbf{n},x} + \text{ h.c.}  \right)
\end{equation}

Now, we have the 2-dimensional Hamiltonian written in terms of only logical operations. Let us define the bosonic operators and the corresponding number operator following the same construction as in Sec.~\ref{section:4} of the main text
\begin{equation}
    \phi_{\mathbf{n}, \mu}^\dag = \sum_{n=0}^{N-2} \sqrt{n+1} \ket{n+1}\bra{n} \quad\quad
    \phi_{\mathbf{n}, \mu} = \sum_{n=0}^{N-2} \sqrt{n+1} \ket{n}\bra{n+1} \quad\quad n_{\mathbf{n}, \mu} = \phi^\dagger_{\mathbf{n}, \mu} \phi_{\mathbf{n}, \mu} = \sum_{n=0}^{N-1} n\ket{n}\bra{n}\;.
\end{equation}
Similarly to what was done in Sec.~\ref{section:4} we also introduce $\rho_{\mathbf{l},\mu} = \QFT n_{\mathbf{l},\mu} \QFT^\dagger$, the Fourier transform of the number operator. This is already enough to write the Hamiltonian in terms of only bosonic degrees of freedom since every logical operation can be written in terms of those. However, the Hamiltonian can be significantly simplified, following the same calculations for the 1 dimensional case. Starting from the purely gauge part we have the same electric term as in one spatial dimension
\begin{equation}
H_E=-2\lambda_E\sum_{\mathbf{n}, \mu} \cos\left( \frac{2\pi}{N} n_{\mathbf{n}, \mu} \right)\;,
\end{equation}
while the plaquette term instead takes the following form
\begin{equation}
H_P=-2\lambda_P\sum_{\mathbf{n}} \cos\left( \frac{2\pi}{N} \left(\rho_{\mathbf{n}, x}+\rho_{\mathbf{n+x}, y}-\rho_{\mathbf{n}, y}-\rho_{\mathbf{n+y}, x}\right) \right)\;.
\end{equation}
For the fermionic matter instead, the mass term becomes
\begin{equation}
\begin{split}
    H_M &= \frac{m}{N} \sum_{\mathbf{n}} (-1)^{n_x+n_y} \sum_{j=0}^{N-1} \omega^{j(-1+p_{\mathbf{n}} -n_{\mathbf{n}-x,x} - n_{\mathbf{n}-y,y} + n_{\mathbf{n},x} + n_{\mathbf{n},y}}) \\
    &= m \sum_{\mathbf{n}} (-1)^{n_x+n_y} \delta_N( n_{\mathbf{n},x} + n_{\mathbf{n},y} -n_{\mathbf{n}-x,x} - n_{\mathbf{n}-y,y} +p_{\mathbf{n}} - 1) \\
\end{split}
\end{equation}
In analogy to what we have obtained for the case of one spatial dimension. As we did there, it is convenient to denote the delta as follows
\begin{equation}
\pi_{\mathbf{n}}=\delta_N( n_{\mathbf{n},x} + n_{\mathbf{n},y} -n_{\mathbf{n}-x,x} - n_{\mathbf{n}-y,y} +p_{\mathbf{n}} - 1)\;,
\end{equation}
so that the mass term can be more conveniently written as
\begin{equation}
H_M = m \sum_{\mathbf{n}} (-1)^{n_x+n_y}\pi_{\mathbf{n}}\;.
\end{equation}
In the same way, we can also write the string of Jordan-Wigner phase factors as
\begin{equation}
\begin{split}
\label{eq:pn_in2dim}
    P_{\mathbf{n}} = &\prod_{\mathbf{l}=\mathbf{n}}^{\mathbf{n}+y} \left(\delta_N( n_{\mathbf{l},x} + n_{\mathbf{l},y} -n_{\mathbf{l}-x,x} - n_{\mathbf{l}-y,y} +p_{\mathbf{l}} - 1)-\delta_N( n_{\mathbf{l},x} + n_{\mathbf{l},y} -n_{\mathbf{l}-x,x} - n_{\mathbf{l}-y,y} +p_{\mathbf{l}}) \right)\\
    &\equiv\prod_{\mathbf{l}=\mathbf{n}}^{\mathbf{n}+y} (-1)^{1-\delta_N( n_{\mathbf{l},x} + n_{\mathbf{l},y} -n_{\mathbf{l}-x,x} - n_{\mathbf{l}-y,y} +p_{\mathbf{l}}-1)}=(-1)^{\sum_{\mathbf{l}=\mathbf{n}}^{\mathbf{n}+y}(1-\pi_{\mathbf{l}})}\;,
\end{split}
\end{equation}
where the second line can be obtained by noticing that only one of the two Kronecker deltas is non-zero in the physical subspace and that all terms commute with each other. The hopping term in the $x$ direction follows in direct analogy with the one dimensional case
\begin{equation}
H_{hop}^x = -\epsilon \sum_{\mathbf{n}}\left(\pi_{\mathbf{n}}e^{i\frac{2\pi}{N}\rho_{\mathbf{n}}}\pi_{\mathbf{n}+x}+\pi_{\mathbf{n}+x}e^{-i\frac{2\pi}{N}\rho_{\mathbf{n}}}\pi_{\mathbf{n}}\right)
\end{equation}
while in the $y$ direction we need to include the Jordan-Wigner strings as
\begin{equation}
H_{hop}^y = -\epsilon \sum_{\mathbf{n}}P_{\mathbf{n}}\left(\pi_{\mathbf{n}}e^{i\frac{2\pi}{N}\rho_{\mathbf{n}}}\pi_{\mathbf{n}+y}+\pi_{\mathbf{n}+y}e^{-i\frac{2\pi}{N}\rho_{\mathbf{n}}}\pi_{\mathbf{n}}\right)\;.
\end{equation}

Note that the identification of the $P_\mathbf{n}$ operators as being composed by purely phases in the second line of Eq.~\eqref{eq:pn_in2dim} ensures that this term can be implemented as a unitary operator. This directly circumvents the need to introduce the unitary operator $\zeta$ in Eq.~\eqref{eq:zeta_JW} of the main text.

\section{Qudits error correction}
\label{Appendix:qudits_error_correction}

This section summarizes the main differences between qubits and qudits when talking about quantum error correction. In particular, we will explain the stabilizer formalism for qudits, and the qudit equivalent Clifford+T universal gate set. Furthermore, we will provide an example with the phase-flip code.

In the qubit case, the Pauli group $P$ is defined as the group generated by the Pauli matrices $P = \langle \mathbb{1}, X, Z, Y \rangle$, and the $n$-qubit Pauli group $P_n = P^{\otimes n}$ is then generated by taking $n$ tensor products of the single qubit Pauli group. From here, one can define the Stabilizer group $S$ as an Abelian subgroup of the Pauli group, and this defines a quantum error correcting code. In particular, if $S$ has $n-k$ generators, we will say that the quantum error correcting code defined by $S$ has parameters $[[n,k,d]]$, where $n$ is the number of physical qubits, $k$ is the number of logical qubits, and $d$ is the distance of the code, which is related to the maximum number of errors that the code is able to correct. In this framework, we also define the normalizer of $S$ as the set of operators that commute with every operator in $S$. Operators that are in the normalizer of $S$ but not in $S$, are called logical operations (non-detectable errors), while detectable errors are operators that are not in the normalizer of $S$ (so they are operators that anticommute with at least one element of $S$).

As a small mathematical remark, in order to simplify the treatment with qudits below, let us define the Pauli group in a slightly different way. The single-qubit Pauli group as we defined it above contains 16 elements:
\begin{equation}
    P = \{ \pm \mathbb{1}, \pm i\mathbb{1}, \pm X, \pm iX, \pm Y, \pm iY, \pm Z, \pm iZ, \}
\end{equation}
However, since $XY=iZ$ and $\mathbb{1} = X^2 = Z^2$, it would be easier to define $P$ as the group generated only by $X$ and $Z$. This would lead to a smaller group, containing the same operators ($\mathbb{1}, X, Y, Z$) but not every combination of operator and phases (indeed, it would contain $\{ \pm\mathbb{1}, \pm X, \pm Z, \pm iY \}$). By taking this fact into account, we have that the two groups are equivalent if we mod out phases:
\begin{equation}
    \langle X, Z \rangle / \{ \pm 1, \pm i \} = P / \{ \pm 1, \pm i \}
\end{equation}
and in general $\langle X, Z \rangle^{\otimes n} / \{ \pm 1, \pm i \} = P_n / \{ \pm 1, \pm i \}$. Since those phases can be always seen as coefficients in front of those matrices, without loss of generality we can define everything modding out phases, and saying that the Pauli group is the group generated by $X$ and $Z$ only.

The stabilizer formalism can be extended in a similar manner from the qubit construction to $N$ levels, assuming the dimensionality of the qudit (the number levels) $N$ to be a prime number. This will be useful while doing error-correction to ensure the group structure to be preserved when going from $2$ to $N$ levels.  Such a generalization is not unique, but here we use the following definition of the generalized Pauli matrices as \cite{Gottesman1999}:
\begin{equation}
\begin{split}
    \Xd \ket{j} &= \ket{(j+1)\mod{N}} ,\quad\quad\quad\quad
    \Zd \ket{j} = \omega^j \ket{j} \;,
\end{split}
    \label{eq:generalized_pauli}
\end{equation}
where $\omega = e^{i2\pi /N}$ is the $N$-th root of unity and $j\in \mathbb{Z}_N$. These matrices can also be expressed in bra-ket notation as
\begin{equation}
\begin{split}
    \Xd &= \sum_{i=0}^{N-1} \ket{(i+1)\mod N}\bra{i} \\
    \Zd &= \sum_{i=0}^{N-1} \omega^i \ket{i}\bra{i} \\
\end{split}
\end{equation}
These generalized Pauli matrices satisfy the following relation
\begin{equation}
    \Xd \Zd = \omega^{-1} \Zd \Xd
\end{equation}
and assuming $N$ to be prime we also have that $N$ is the smallest power such that $\Xd^N = \Zd^N = \mathbb{1}$. These two matrices generate the generalized Pauli group (ignoring phases, as discussed before), which means that every element of the single-qudit Pauli group has the form $\Xd^r \Zd^s$ where $r,s \in \mathbb{Z}_N$. This means that the commutation relation between two operators in the generalized Pauli group is:
\begin{equation}
    (\Xd^r \Zd^s) (\Xd^t \Zd^u) = \omega^{st-ru} (\Xd^t \Zd^u) (\Xd^r \Zd^s)
\end{equation}
One can prove that every element of the generalized Pauli group different from the identity will have eigenvalues $1,\omega, \cdots, \omega^{N-1}$, every one with multiplicity one. It follows trivially from this that such operators are traceless. From now on, when we will speak about multi-qudit systems, we will denote with a subscript the qudit the operator acts on, assuming that it acts with the identity on every other qudit (i.e. $\Xd_i = \mathbb{1}\otimes \cdots \Xd \cdots, \otimes \mathbb{1}$, where $\Xd$ acts on the $i$-th qudit).

We can define the stabilizer group $S$ of a code, like in the qubit case, as an Abelian subgroup of the generalized Pauli group. The code space is then the set of states that are eigenvectors of all elements of $S$ with eigenvalue $+1$. Analogously to the qubit case, if the stabilizer code has $n$ physical qudits and $n-k$ stabilizers (generators of $S$), then it can encode $k$ logical qudits.  This follows from dimension counting because all stabilizers measure to $+1$ in the code space which limits the dimension of the code space to $N^n/N^{n-k}=N^k$, which restricts us to an $N$ logical qudit space. 

The codespace $C$ (states protected by the code), is formally defined as a Hilbert space satisfying $3$ conditions \cite{Gheorghiu_2014}:
\begin{enumerate}
    \item There is a subgroup $S$ of the $n$-qudit Pauli group such that every element of $C$ is eigenvector of every element of $S$ with eigenvalue $+1$
    \item $S$ is maximal in the sense that every operator in the Pauli group that satisfies the first condition is in $S$
    \item $C$ is maximal in the sense that every ket that satisfies the first condition is in $C$
\end{enumerate}

In the binary case, detectable errors are those operators that anti-commute with at least one element of the stabilizer. Here we say that we can detect an error $E$ if it is an operator that satisfies $EM = \omega^a ME$ with $a\ne 0 \mod N$ for at least $1$ operator $M$ in the stabilizer $S$. In this way, the state $E\ket{\psi}$ will give eigenvalue $\omega^a$ instead of $+1$ and we can detect the error by measuring the eigenvalue of $M$. If we have $n$ physical qudits, and the stabilizer $S$ is a group with $n-k$ generators, we will denote the code as $[[n,k,d]]_N$. The fact the number of logical qudits is $k$ follows directly from the fact that operators in the Pauli group that are not the identity have one eigenvalue equal to $1$, with multiplicity $1$. This means that every time we add a generator to $S$, we divide the number of codewords in $C$ by $N$ (this holds only if $N$ is a prime number, as we assumed before). 

The subgroup $S$ can be extended to a set of $n$ commuting operators by choosing $k$ additional operators $\bar \Zd_1, \cdots, \bar \Zd_k$. These operators will be the logical $\Zd$ operations on the $k$ encoded qudits. Then, we can define the logical $\Xd$ operations on the $k$ encoded qudits by defining $k$ operators $\bar \Xd_1, \cdots, \bar \Xd_k$ that commutes with every $M\in S$ and satisfy the following relations:
\begin{equation}
\begin{split}
    \bar \Xd_i \bar \Zd_j &= \bar \Zd_j \bar \Xd_i ~~~~~~ i \ne j \\
    \bar \Xd_i \bar \Zd_i &= \omega^{-1} \bar \Zd_i \bar \Xd_i
\end{split}
\end{equation}
These $n+k$ operators generate the group of all Pauli operators that commute with $S$, which means that they generate the normalizer of $S$. As in the binary case, operators that are in the normalizer are either in $S$, and so act trivially on the codewords, or are not in $S$ and so are errors that our code cannot detect (the logical operations). The operators in the generalized Pauli group that are outside the normalizer are detectable errors.

The distance $d$ of the code is defined, both in the qubit and qudit case, as the minimum Hamming weight of the logical operations, which in other words means the minimum weight that an operator has to have to be undetectable. The maximum number of errors that the code is able to correct is then $\lfloor (d-1)/2 \rfloor$.

We will now assume the error model to be a generalized Pauli channel where the probability of an arror $\Zd^j$ or $\Xd^j$ to happen is the same for every $j \ne 0 \mod N$ (this simply means that we are considering the error $\Xd$ and $\Xd^2$ to have the same probability to occur).

It is useful to see that we can recover the symmetry between the $\Xd$ and $\Zd$ bases by defining the generalization of the Hadamard gate on a qudit as the gate that conjugates $\Xd$ and $\Zd$. This gate is simply the quantum Fourier transform defined as follows:
\begin{equation}
\begin{split}
%    \QFT \ket{j} &= \frac{1}{\sqrt{N}} \sum_{k=0}^{N-1} \omega^{jk} \ket{k} \\
    \QFT &= \frac{1}{\sqrt{N}} \sum_{k=0}^{N-1} \sum_{j=0}^{N-1} \omega^{jk} \ket{k}\bra{j} \\
\end{split}
\end{equation}
Then, we have the usual relation $\QFT\ \Zd\ \QFT^\dag = \Xd$.

\subsection{Gates and measurements}
\label{sec:gates_and_measurements}

We already defined the generalization of Pauli gates and the generalization of the Hadamard gate (which corresponded to the $\QFT$), but these are not enough to generate a universal set of gates. So, first, let us define the generators for a universal single-qudit gate set as the single qudit Clifford group plus one non-Clifford gate. As in the qubit case, we can define the Clifford group as the group that maps the generalized Pauli group of $n$ qudits $P_n$ in itself (apart a phase $\omega^k$ for some integer $k$):
\begin{equation}
    \mathcal{C} = \left\{ U\in SU(d) : UMU^\dag \in \omega^k P_n, \forall M \in P_n \right\}
\end{equation}

Following \cite{jain2020normalformsinglequditcliffordt}, we can define the $\Sd$ gate as follows:
\begin{equation}
    \Sd = \sum_{j=0}^{N-1} \omega^{j(j+1)[2^{-1}]_N} \ket{j}\bra{j}
\end{equation}
where $[2^{-1}]_N$ is the inverse of $2$ in the $\mathbb{Z}_N$ cyclic group (for example, for $N=5$, $[2^{-1}]_5=3$). One can prove that the single-qudit Clifford group is generated by $\Sd$ and $\QFT$. In Appendix \ref{App:cliffordST} we provide the proof that $\Sd$ as we defined it is Clifford, while one can find the proof that those operators generate the entire qudit Clifford group in \cite{jain2020normalformsinglequditcliffordt, Wang_2020, GLAUDELL201954}. Then, as in the qubit case, we need only one additional operator, which is not Clifford, in order to generate a universal set of single-qudit gates. In this paper we will consider the following definition of qudit $\Td$ gate:
\begin{equation}
    \Td = \sum_{j=0}^{N-1} \omega^{j^3 [6^{-1}]_N} \ket{j}\bra{j}
\end{equation}
We provide the proof that $\Td$ is not Clifford in Appendix \ref{App:cliffordST}. This definition of $\Td$ gate only works for $N\ge 5$, and since we are considering $N$ to be a prime number, it means that the only non-qubit case for which it does not work is $N=3$. In the qutrit case, we can define the $\Td$ gate as follows:
\begin{equation}
    \Td = \begin{pmatrix}
        e^{i\frac{2\pi}{9}} & 0 & 0 \\
        0 & 1 & 0 \\
        0 & 0 & e^{-i\frac{2\pi}{9}} \\
    \end{pmatrix}
\end{equation}
Those definitions of $\Td$ gate are equally valid for the universality. However, in the case of $N\ge 5$ we have an additional property that we do not have in the qutrit case. This property is $\Td \Xd \Td^\dagger$ being Clifford for $N\ge 5$, and this will be important when discussing state injection protocols in Section \ref{section:universal_gate_set}. For all other purposes, the definition for $N=3$ is equivalent to the one for $N\ge 5$.

\smallskip

Qudit operations can further be easily generalized to a multi-qudit system. Indeed, one can consider the universal single-qudit gate set, and simply add an entangling gate. We can define the control-$\Zd$ gate as follows:
\begin{equation}
    C\Zd = \sum_{j=0}^{N-1} \ket{j}\bra{j}\otimes \Zd^j = \sum_{j,k=0}^{N-1} \omega^{jk} \ket{j}\bra{j} \otimes \ket{k}\bra{k}
\end{equation}
This is an entangling gate that, together with the single-qudit Clifford group generates the multi-qudit Clifford group, while together with the single-qudit universal gate-set generates a multi-qudit universal gate-set.

Starting from this gate, we can define two useful 2-qudit gates: the $\SUMgate$ and the $\widetilde{C\Xd}$ gates. Their definition is as follows:
\begin{equation}
\begin{split}
    \SUMgate &= (\mathbb{1}\otimes \QFT)C\Zd(\mathbb{1}\otimes \QFT^\dag) \\
    \widetilde{C\Xd} &= (\mathbb{1}\otimes \QFT) C\Zd (\mathbb{1}\otimes \QFT)
\end{split}
\end{equation}
Both these gates are a possible generalization of the qubit CNOTs to $N$-level systems. Let us start by analyzing the $\SUMgate$ gate. It is easy to see that:
\begin{equation}
\begin{split}
    \SUMgate \ket{j}\ket{k} &= \ket{j}\ket{(k+j) \mod N} \\
    \SUMgate &= \sum_{i=0}^{N-1} \ket{i}\bra{i}\otimes \Xd^{i}
\end{split}
\end{equation}
This gate is very useful in quantum error correction since it can be easily used to measure the parity between qudits. Consider 2 qubits, and the stabilizer $XX$. In order to measure this stabilizer, one could run a circuit with two CNOTs controlled from the two qubits and targeting an ancilla in the $\ket{0}$ state. In this way, the ancilla after the CNOTs would be in the $\ket{0}$ state if and only if the two qubits were in the same state. In the qudit case, we can generalize the same circuit. The final state in the ancilla should be the difference of the states of the two qudits, so that it would result $\ket{0}$ if and only if the two qudits are in the same state. By substituting the CNOTs with the $\SUMgate$ and inverting one of the two gates, we get the following circuit:
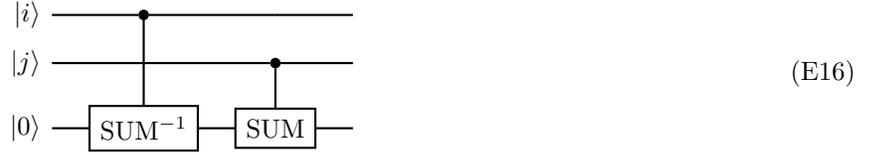
\begin{equation}
\begin{quantikz}
    \lstick{$\ket{i}$} & \ctrl{2} & \qw & \qw \\
    \lstick{$\ket{j}$} & \qw & \ctrl{1} & \qw \\
    \lstick{$\ket{0}$} & \gate{\SUMgate^{-1}} & \gate{\SUMgate} & \qw 
\end{quantikz}
\end{equation}
the final state on the ancilla will be $\ket{j-i}$, which was what we wanted.

The other possible generalization of the CNOT, is what we called $\widetilde{C\Xd}$:
\begin{equation}
    \widetilde{C\Xd}\ket{j}\ket{k} = \ket{j}\ket{(-k-j) \mod N}
    \label{eq:CX_tilde}
\end{equation}
The $\SUMgate$ gate is useful also because it is very intuitive and easy to use, like in the case of the parity check measurement. The $\widetilde{C\Xd}$ gate instead, is useful because it simplifies some constructions like the one of the SWAP gate. Indeed, to perform a two-qudit SWAP operation with the $\SUMgate$ gate, we need 3 $\SUMgate$ and an auxiliary gate $K$ that performs $K\ket{j}=\ket{-j}$. By using the $\widetilde{C\Xd}$ gate instead, we need no single qudit gates as in the qubits case. Moreover, note that $\widetilde{C\Xd} = \widetilde{C\Xd}^\dag$ \cite{Wang_2020}.

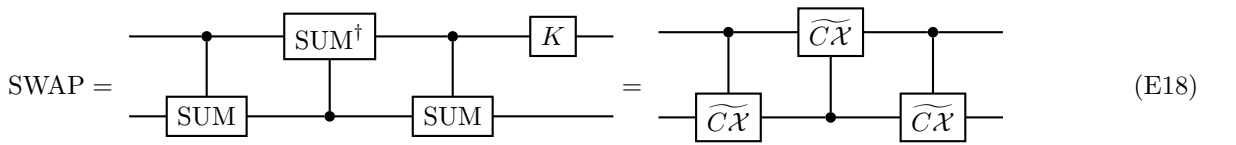
\begin{equation}
\text{SWAP} = 
\begin{quantikz}
    & \ctrl{1} & \gate{\SUMgate^\dag} & \ctrl{1} & \gate{K} & \qw \\
    & \gate{\SUMgate} & \ctrl{-1} & \gate{\SUMgate} & \qw & \qw \\
\end{quantikz}
=
\begin{quantikz}
    & \ctrl{1} & \gate{\widetilde{C\Xd}} & \ctrl{1} & \qw & \\
    & \gate{\widetilde{C\Xd}} & \ctrl{-1} & \gate{\widetilde{C\Xd}} & \qw \\
\end{quantikz}
\end{equation}

\subsection{Qudit phase-flip code}
\label{sec:qudit_phase_flip_code}

We provide an example of error-correcting code for qudits in this section: the phase-flip code. This code is able to correct one single-qubit $\Zd$ error, which means that, with our error model, it is able to correct $\Zd^j$ for every $j$.

In order to build a phase-flip code, consider the simple case of a $3$-level system. The generalized Pauli matrices for qutrits are:
\begin{equation}
    \Xd = \left( \begin{matrix}
        0 & 0 & 1 \\
        1 & 0 & 0 \\
        0 & 1 & 0 \\
    \end{matrix} \right)
    ~~~~~~~~
    \Zd = \left( \begin{matrix}
        1 & 0 & 0 \\
        0 & \omega & 0 \\
        0 & 0 & \omega^2 \\
    \end{matrix} \right)
\end{equation}
where $\omega = e^{i2\pi /3}$ is the third root of unity. In general, every single qudit error can be written as a superposition of $\Xd, \Xd^2, \Zd, \Zd^2$ (remember that $\Xd^2 = \Xd^{-1}$).

Consider $3$ physical qutrits that encode $1$ logical qutrit, and the following $2$ stabilizers:
\begin{equation}
\begin{split}
    S_1 &= \Xd_1 \Xd_2^{-1} \\
    S_2 &= \Xd_2 \Xd_3^{-1} \\
\end{split}
\label{eq:stabilizers_phase_flip}
\end{equation}
The logical operations follow from the stabilizers:
\begin{equation}
\begin{split}
    \bar \Zd &= \Zd_1\Zd_2\Zd_3\\
    \bar \Xd &= \Xd_1 = \Xd_2 = \Xd_3 \\
\end{split}
\end{equation}
Looking at Table \ref{tab:error_syndrome_phase-flip_z3}, one can see that we can identify every possible phase error and correct it. As for the codewords, we can define them as one usually does for qubits:
\begin{equation}
\begin{split}
    \ket{0}_L &= \frac{1}{\sqrt{3^2}}\prod_{i=1}^2 (1 + S_i + S_i^2) \ket{000} \\
    \ket{1}_L &= \bar \Xd \ket{0} \\
    \ket{2}_L &= \bar \Xd^2 \ket{0} \\
\end{split}
\end{equation}
This, in our particular case, would lead to the following state:
\begin{equation}
\begin{split}
    \ket{0}_L = \frac{1}{3} (&\ket{000} + \ket{021} + \ket{012} + \ket{120} + \\
    &\ket{210} + \ket{111} + \ket{102} + \ket{201} + \ket{222}) \\
\end{split}
\end{equation}

\begin{table}[h]
    \centering
    \begin{tabular}{|c|c|c|}
    \hline
        $S_1$ & $S_2$ & error \\
    \hline
        $\omega^2$ & $1$ & $\Zd_1$ \\
    \hline
        $\omega$ & $1$ & $\Zd_1^2$ \\
    \hline
        $\omega$ & $\omega^2$ & $\Zd_2$ \\
    \hline
        $\omega^2$ & $\omega$ & $\Zd_2^2$ \\
    \hline
        $1$ & $\omega$ & $\Zd_3$ \\
    \hline
        $1$ & $\omega^2$ & $\Zd_3^2$ \\
    \hline
    \end{tabular}
    \caption{Error syndrome for the phase-flip repetition code. The stabilizers are in Eq.~\eqref{eq:stabilizers_phase_flip}.}
    \label{tab:error_syndrome_phase-flip_z3}
\end{table}

The generalization of the qutrit case to a $N$-level system is straightforward.   The stabilizers and the logical operations remain the same, while the logical states can be written as follows:
\begin{equation}
\begin{split}
    \ket{0} &= \frac{1}{\sqrt{N^2}} \prod_{i=1}^2 \sum_{j=0}^{N-1} (S_i)^j \ket{000} \\
    \ket{n} &= \bar \Xd^n \ket{0} \\
\end{split}
\end{equation}
For a generic code with $k$ stabilizers, the product becomes a product of $k$ terms with $i=1,\cdots,k$ and the normalization factor will be $1/\sqrt{N^k}$.

 Measurement of stabilizers is straightforward to generalize the qubit case. The Hadamard gate becomes the quantum Fourier transform, and the CNOTs become $\SUMgate$ gates. So, to measure the stabilizer $S_1 = \Xd_1 \Xd_2^{-1}$ and $S_2 = \Xd_2 \Xd_3^{-1}$, we can use the circuit depicted in figure \ref{fig:stabilizer_measurement_phase-flip_Zp}. After the measurements of the two ancilla qudits, we will need to correct the error accordingly, by applying $\Zd$ or $\Zd^2$ where needed (to correct a $\Zd$ error we have to apply $\Zd^2$ and vice versa).

\begin{figure}[h]
    \centering
    \begin{quantikz}[row sep={0.8cm,between origins},column sep={0.20cm}]
    & \gate{{\rm QFT}} & \ctrl{3} & \qw & \qw & \gate{{\rm QFT}^\dag} & \qw \\
    & \gate{{\rm QFT}} & \qw & \ctrl{2} & \qw & \gate{{\rm QFT}^\dag} & \qw \\
    & \gate{{\rm QFT}} & \qw & \qw & \ctrl{2} & \gate{{\rm QFT}^\dag} & \qw \\
    \lstick{$\ket{0}$} & \qw & \gate{\text{SUM}} & \gate{\text{SUM}^{-1}} & \qw & \meter{} \\
    \lstick{$\ket{0}$} & \qw & \qw & \gate{\text{SUM}}\vqw{-1} & \gate{\text{SUM}^{-1}} & \meter{} \\
    \end{quantikz}
    \caption{Quantum circuit to measure error syndromes in the phase-flip code. The first ancilla will measure the stabilizer $S_1 = \Xd_1 \Xd_2^{-1}$, while the second ancilla will measure $S_2 = \Xd_2 \Xd_3^{-1}$.}
    \label{fig:stabilizer_measurement_phase-flip_Zp}
\end{figure}

This works due to the fact that the $\SUMgate$ gate has the same error-propagation as the CNOT in the qubit case, as shown in Equations \ref{eq:X_error_propagation} and \ref{eq:Z_error_propagation}.

\begin{equation}
\begin{quantikz}[row sep={0.8cm,between origins},column sep={0.25cm}]
    & \gate{\Xd} & \ctrl{1} & \qw \\
    & \qw & \gate{\SUMgate} & \qw \\
\end{quantikz}
=
\begin{quantikz}[row sep={0.8cm,between origins},column sep={0.25cm}]
    & \ctrl{1} & \gate{\Xd} & \qw \\
    & \gate{\SUMgate} & \gate{\Xd} & \qw \\
\end{quantikz}
\label{eq:X_error_propagation}
\end{equation}
\begin{equation}
\begin{quantikz}[row sep={0.8cm,between origins},column sep={0.25cm}]
    & \qw & \ctrl{1} & \qw \\
    & \gate{\Zd} & \gate{\SUMgate} & \qw \\
\end{quantikz}
=
\begin{quantikz}[row sep={0.8cm,between origins},column sep={0.25cm}]
    & \ctrl{1} & \gate{\Zd} & \qw \\
    & \gate{\SUMgate} & \gate{\Zd} & \qw \\
\end{quantikz}
\label{eq:Z_error_propagation}
\end{equation}

The relation in Eq.~\eqref{eq:inversion_SUM_gate} shows how target and control qudits in a $\SUMgate$ gate can be exchanged by using the $\QFT$.

\begin{equation}
\begin{quantikz}[row sep={0.8cm,between origins},column sep={0.25cm}]
    & \gate{\QFT} & \ctrl{1} & \gate{\QFT^\dag} & \qw \\
    & \gate{\QFT} & \gate{\SUMgate} & \gate{\QFT^\dag} & \qw \\
\end{quantikz}
= 
\begin{quantikz}[row sep={0.8cm,between origins},column sep={0.25cm}]
    & \gate{\SUMgate} & \qw \\
    & \ctrl{-1} & \qw \\
\end{quantikz}
\label{eq:inversion_SUM_gate}
\end{equation}

Furthermore, we can use the $\SUMgate$ gate as a CNOT also for flag qudits, if we take into account the fact that $\text{SUM}\ne \text{SUM}^{-1}$. This means that we can add flag qudits if needed, for example, to measure a higher weighted stabilizer, as shown in Figure \ref{fig:flag_z3_example} \cite{Durso-Sabina}.

\begin{figure}[h]
    \centering
    \resizebox{0.47\textwidth}{!}{
    \begin{quantikz}[row sep={0.8cm,between origins},column sep={0.15cm}]
    \lstick{$\ket{a}$} & \ctrl{4} & \qw  & \qw & \qw & \qw & \qw & \qw \\
    \lstick{$\ket{b}$} & \qw & \qw  & \ctrl{3} & \qw & \qw & \qw & \qw \\
    \lstick{$\ket{c}$} & \qw & \qw  & \qw & \ctrl{2} & \qw & \qw & \qw \\
    \lstick{$\ket{d}$} & \qw & \qw  & \qw & \qw & \qw & \ctrl{1} & \qw \\
    \lstick{$\ket{0}$} & \gate{\SUMgate} & \gate{\SUMgate}  & \gate{\SUMgate} & \gate{\SUMgate} & \gate{\SUMgate^{-1}} & \gate{\SUMgate} & \qw \\
    & & \ctrl{-1}  & \qw & \qw & \ctrl{-1} & \meter{} & \\
    \end{quantikz}}
    \caption{Example circuit to show the use of flag qudits to identify $\Zd$ errors to deduce their propagation through a circuit as explained in \cite{Durso-Sabina}.}
    \label{fig:flag_z3_example}
\end{figure}

\bibliography{biblio,biblio_staggered}

\end{document}